%% file: T2K_H20_Pion.tex
\documentclass[
reprint,
superscriptaddress,
showpacs,
nofootinbib,
nobibnotes,
 amsmath,amssymb,
 aps,
prd,
]{revtex4-1}
\usepackage{graphicx}
\usepackage{dcolumn}
\usepackage{bm}
\usepackage{hyperref}
\begin{document}

\title{
Modeling neutrino-induced charged pion production on water at T2K kinematics}

\author{A. Nikolakopoulos}
\email{alexis.nikolakopoulos@ugent.be}
\affiliation{Department of Physics and Astronomy, Ghent University \\
Proeftuinstraat 86, B-9000 Gent, Belgium}
\author{R. Gonz\'{a}lez-Jim\'{e}nez}
\email{raugonjim@gmail.com}
\affiliation{Department of Physics and Astronomy, Ghent University \\
Proeftuinstraat 86, B-9000 Gent, Belgium}
\author{K. Niewczas}
\affiliation{Department of Physics and Astronomy, Ghent University \\
Proeftuinstraat 86, B-9000 Gent, Belgium}
\affiliation{Institute of Theoretical Physics, University of Wroc{\l}aw \\
Plac Maxa Borna 9, 50-204 Wroc{\l}aw, Poland
}
\author{J. Sobczyk}
\affiliation{Institute of Theoretical Physics, University of Wroc{\l}aw \\
Plac Maxa Borna 9, 50-204 Wroc{\l}aw, Poland
}
\author{N. Jachowicz}
\email{natalie.jachowicz@ugent.be}
\affiliation{Department of Physics and Astronomy, Ghent University \\
Proeftuinstraat 86, B-9000 Gent, Belgium}
\date{\today}

\begin{abstract}
\begin{description}
\item[Background]
Pion production is a significant component of the signal in accelerator-based neutrino experiments.
Over the last years, the MiniBooNE, T2K and MINERvA collaborations have reported a substantial amount of data on (anti)neutrino-induced pion production on the nucleus.
However, a comprehensive and consistent description of the whole data set is still missing. 
\item[Purpose]
We aim at improving the current understanding of neutrino-induced pion production on the nucleus.
To this end, the comparison of experimental data with theoretical predictions, preferably based on microscopic models, is essential to disentangle the different reaction mechanisms involved in the process. 
\item[Method]
To describe single-pion production (SPP) we use a hybrid model that combines a low- and a high-energy approach.
The low-energy model (LEM) contains resonances and background terms. At high invariant masses, a high-energy model based on a Regge approach is employed.
The model is implemented in the nucleus using the relativistic plane wave impulse approximation (RPWIA).
\item[Results]
We present a comparison of the hybrid-RPWIA and LEM with the recent neutrino-induced charged current $1\pi^+$ production cross section on water reported by T2K.
In order to judge the impact of final-state interactions (FSI) we confront our results with those of the NuWro Monte Carlo generator.
\item[Conclusions]
The hybrid-RPWIA model and NuWro compare favorably to the data, albeit that FSI are not included in the former.
The need of a high-energy model at T2K kinematics is made clear.
These results complement our previous work [Phys. Rev. D 97, 013004 (2018)] where we compared the models to the MINERvA and MiniBooNE $1\pi^+$ data.
The hybrid-RPWIA model tends to overpredict both the T2K and MINERvA data in kinematic regions where the largest suppression due to FSI is expected, 
 and agrees remarkably well with the data in other kinematic regions. On the contrary, the MiniBooNE data is underpredicted over the whole kinematic range.

\end{description}
\end{abstract}

\pacs{25.30.Pt, 12.15.-y, 13.15.+g, 13.60.Le}
\maketitle


\section{\label{sec:introduction}Introduction}
Neutrino energies from beams in accelerator-based experiments, such as MiniBooNE~\cite{MB:pion,MB:CCneutralpion}, T2K~\cite{T2K:Inclnumu,T2KCC1PIH2O}, MINERvA~\cite{MINERvA:CC1PI,MINERvA:CCPI0} and NOvA~\cite{NOVA},
are spread over a broad range with contributions from increasingly more energetic neutrinos (as is the case in e.g. DUNE~\cite{DUNE}).
As the energy of the incoming neutrino in an interaction is not precisely known, all measurements are averaged over the incoming neutrino flux.
This means that the interaction of the neutrino with nuclear targets should be known and reliably described over a large energy range in order to be able to extract neutrino mixing parameters~\cite{NUSTECWP}. 
Single pion production (SPP) provides a significant contribution to the signal in current and future oscillation experiments.
In addition to this, neutrino-induced pion production is important in unraveling the axial structure of the nucleon.

In this paper we compare the predictions of the hybrid relativistic plane wave impulse approximation (hybrid-RPWIA) model for SPP with the charged-current single charged pion (CC1$\pi^+$) cross section on water reported by the T2K experiment \cite{T2KCC1PIH2O}.
The T2K $\nu_{\mu}$-flux has a peak for neutrino energies of approximately $600~\mathrm{MeV}$.
The CC1$\pi^+$ signal in this energy region mostly consists of elementary single-pion production through the decay of the delta resonance. 
The delta region is the main focus of most models describing SPP~\cite{MartiniModel:2009,ReinSeghal,Hernandez:Pion,Paschos:2015,SUSA:2016,BussMosel,Praet,Singh:2006,Singh:2016,ZhangSerot,Mosel:MB,Sato,Hernandez:PionNucleus}.
Most models that aim at describing the low energy resonance region tend to exhibit problematic behavior when they are extended to large values of invariant mass ($W \gtrsim 1.4~\mathrm{GeV}$) because only first-order diagrams are taken into account \cite{Gonzalez:SPPnucleon}.
As an exception we mention the coupled channel model of Nakamura et al.~\cite{Sato}, which can be extended to larger values of invariant mass ($W\lesssim2~\mathrm{GeV}$) through unitarization of the amplitudes.

The hybrid model for SPP on the nucleon is described in Ref.~\cite{Gonzalez:SPPnucleon}. 
The aim is to describe the elementary reaction over a large range of the invariant mass.
The formalism is based on the combination of the first order background diagrams obtained from the Chiral Perturbation Theory (ChPT) Lagrangian density for the $\pi N$-system \cite{ChPT}, with the contributions of the delta and more massive isospin-1/2 resonances [$P_{11}(1440)$, $S_{11}(1535)$, and $D_{13}(1520)$] \cite{Hernandez:Pion,Hernandez:PionNucleus,Lalakulich:Res}.
For the resonances, the s- and u-channel diagrams are included. 
The resonant amplitudes are regularized by a Gaussian-dipole form factor \cite{FFres,Vrancx} in order to retain the correct amplitude when $s(u)\approx M^2_{res}$, meanwhile eliminating the unphysical 
contributions far away from the resonance peak.
For high values of the invariant mass, the non-resonant amplitudes present pathologies due to the fact that only the lowest order diagrams are considered \cite{Gonzalez:SPPnucleon}. 
Taking into account higher order diagrams quickly becomes unfeasible. Alternatively, the high-energy region can be readily described by a Regge approach, which provides the correct $s$-dependence of the amplitude at high $W$.
Our approach is based on the procedure for ``reggeizing'' the non-resonant background as proposed in Refs.~\cite{GVL,Kaskulov:Mosel} for the vector current contributions, which was extended to the axial current in Ref. \cite{Gonzalez:SPPnucleon}. 
The low- and high-energy models for the non-resonant contributions are combined by a smooth $W$-dependent transition.

The hybrid model is embedded in the nucleus using the relativistic plane wave impulse approximation (RPWIA).
The hybrid-RPWIA model is described in Ref.~\cite{HybridRPWIA}, and was compared to pion production data presented by MINERvA \cite{MINERvA:CCPI0,MINERvA:CC1PI} and MiniBooNE \cite{MB:CCneutralpion,MB:pion}.
The impulse approximation (IA) is adopted in the sense that we treat the hadronic current as the incoherent sum of single nucleon interactions.
The bound nucleons are modeled by relativistic mean field (RMF) \cite{RINGRMF,WALECKARMF} wavefunctions occupying discrete shells with well-defined angular momentum and binding energy.
The hadronic current in the RPWIA is then obtained by describing the final-state pion and nucleon by plane waves with well-defined momentum. 

The hybrid-RPWIA model is fully relativistic in both the operators and the wavefunctions.
However it does not contain any final state interactions (FSI).
The elastic distortion of the outgoing nucleon and pion is ignored as they are described by plane waves. 
This can be treated consistently in our relativistic quantum mechanical framework by distortion of the outgoing wavefunctions, which will be the next step in this project. 
Inelastic FSI, which should be taken into account to fully describe $\pi^+$ production on the nucleus include pion-absorption, charge-exchange reactions, and secondary pion production.
These processes are usually treated in Monte Carlo generators using intra-nuclear cascade models \cite{Hayato:NEUT,GENIE,NuWroFSI}, or kinetic transport theory \cite{GiBUU}, to propagate the particles originating from an elementary vertex through the nucleus.
As mentioned, the use of reliable microscopic models is essential to gauge the understanding of the fundamental process.
In this work, the effect of FSI is judged by comparing the results to the predictions of the NuWro Monte Carlo event generator, with and without FSI~\cite{NuWroFSI}.

The structure of this paper is as follows. In Sec. \ref{sec:results}, we compare the results of the model to experimental data. 
In the first subsection \ref{sec:results1} the effects of higher mass resonances, and the differences between the LEM and hybrid-RPWIA model are explored.
In subsection \ref{sec:results2} we compare our results with those of the NuWro generator.
 The conclusions are presented in Sec. \ref{sec:conclusion}.
\section{\label{sec:results}Results}
The data was obtained with the ND280 detector in the T2K experiment. The phase space is restricted to $P_{\mu} > 200~\mathrm{MeV}$, $P_{\pi} > 200~\mathrm{MeV}$, $\cos\theta_{\pi}> 0.3$, and $\cos\theta_{\mu} > 0.3$ \cite{T2KCC1PIH2O}.
The signal is defined as a single $\pi^+$ and muon in the final state with no other mesons.
We compute the cross section on the water target by adding the contributions of two free protons, and the nucleons in ${}^{16}$O described within the RMF model. 

\subsection{\label{sec:results1}The hybrid-RPWIA model}
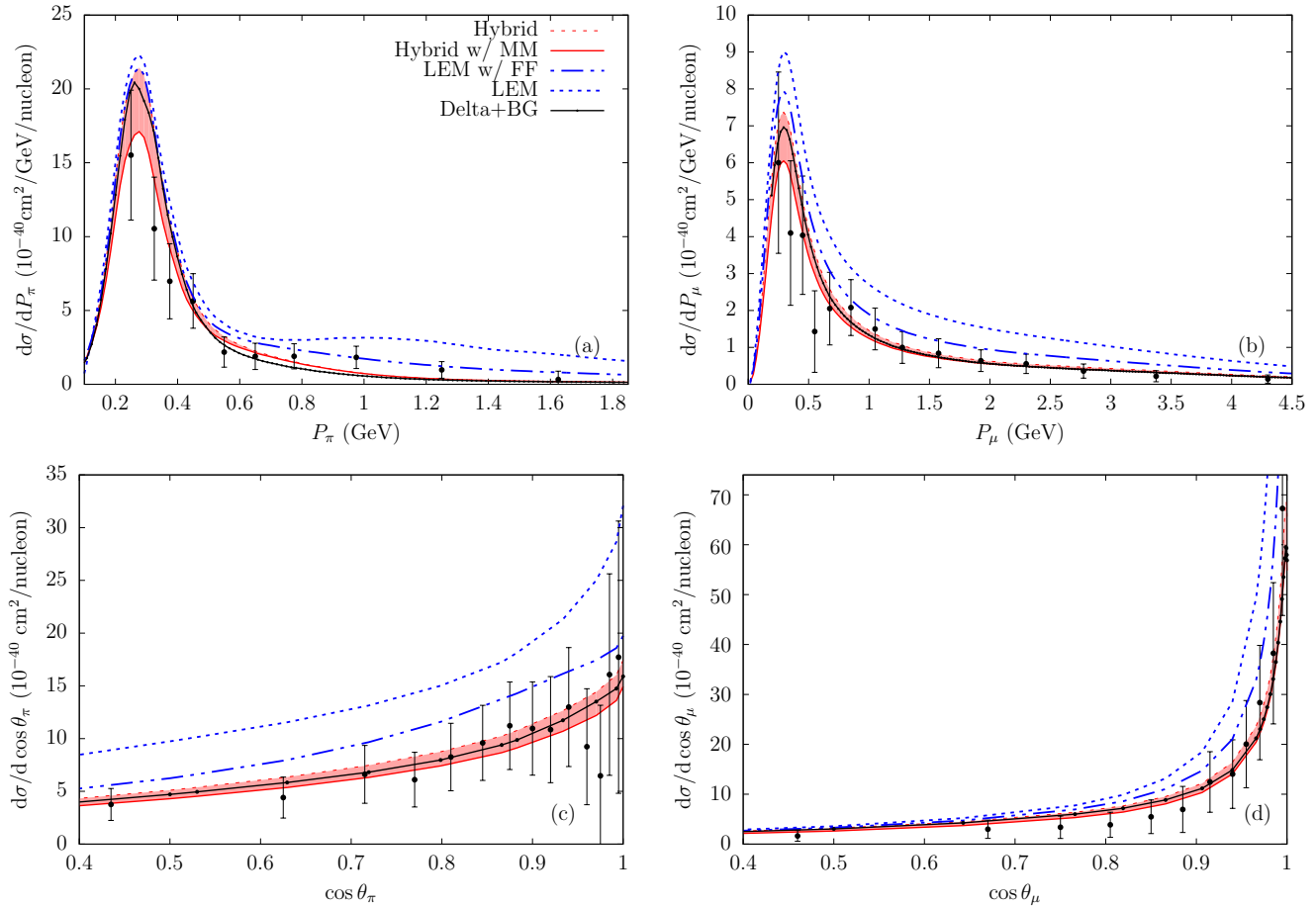
\begin{figure*}
\begin{large}
\resizebox{0.49\textwidth}{!}{\input{./gfx/CS_Ppi.tex}}
\resizebox{0.49\textwidth}{!}{\input{./gfx/CS_Pmu.tex}}
\resizebox{0.49\textwidth}{!}{\input{./gfx/CS_THpi.tex}}
\resizebox{0.49\textwidth}{!}{\input{./gfx/CS_THl.tex}}
\end{large}
\caption{Single differential cross sections for the T2K CC1$\pi^+$ data sample \cite{T2KCC1PIH2O}.
We show the hybrid-RPWIA prediction with and without OSMM of the delta width (dashed and solid red lines respectively), compared to the low energy model (LEM), which consists of the resonant and ChPT background diagrams extended to arbitrarily large values of invariant mass $W$.
The LEM with cutoff form factors for the resonances is depicted with the blue dash-dotted line.
To show the influence of higher mass resonances the hybrid model calculation including only the background and delta resonance is also shown (Delta+BG), it is computed without medium modification of the delta width.}
\label{fig:CS_Ppi}
\end{figure*}

The hybrid-RPWIA model is confronted with the T2K CC1$\pi^+$ data in Fig.~\ref{fig:CS_Ppi}.
Most information on the underlying pion-production mechanism is obtained from the $P_{\pi}$ distribution, Fig.~\ref{fig:CS_Ppi}(a).
The low momentum region around the peak is dominated by delta-mediated pion production, the higher mass resonances are seen to contribute up to around $1~\mathrm{GeV}$, and the Regge approach mainly affects
the high momentum tail.  
The cross section does not heavily depend on $\cos\theta_{\pi}$ as shown in Fig.~\ref{fig:CS_Ppi}(c).
The exact shape at forward scattering angles is obscured by large experimental errorbars.
The dominance of the delta resonance in charged pion production is clear from comparison with the calculation where only the delta and ChPT terms are taken into account, omitting higher mass resonances.
This is labeled as ``Delta+BG'' in Fig. \ref{fig:CS_Ppi}, and is computed without medium modification of the delta width.
The other resonances have isospin $1/2$; therefore, they can only contribute in the $u$-channel to $\pi^+$ production on the proton.
Indeed, the $p(I_3=1/2)+\pi^+(I_3=1)$ final state can only couple to $I_3=+3/2$, allowing no $I=1/2$ resonances in the direct channel.
For a full list of isospin coefficients for the different reaction channels in the hybrid-RPWIA model see for instance table I in Ref.~\cite{Gonzalez:SPPnucleon}.
The influence of the isospin $1/2$ resonances is thus mainly important for interactions with the neutron, where they contribute in the $s$-channel.
In Fig. \ref{fig:CS_Ppi}, we see that the higher mass resonances, contribute up to 20\% of the cross section for $P_{\pi}$ between $0.5~\mathrm{GeV}$ and $1~\mathrm{GeV}$.

Because of the importance of the delta resonance, the medium modification of its decay width leads to a significant suppression of the cross section. 
The width of the delta resonance is modified by the complex part of the delta self-energy in the nuclear medium.
We compute this effect within the Oset and Salcedo medium modification (OSMM) formalism \cite{OSMM,Mosel:MB,Hernandez:PionNucleus}. 
The hybrid-RPWIA model with medium modification of the delta is plotted with the solid red line in Fig.~\ref{fig:CS_Ppi}. 
The uncertainties and inconsistencies pertaining to the use of this procedure for the medium modification of the delta in the framework of our model were discussed in \cite{HybridRPWIA}.
In particular the $\Delta N \rightarrow \pi N$ process is included in the modification of the width, a process that contributes to the experimental signal.
The contribution of this channel has previously been modeled by multiplying the delta amplitude by a weighting factor, which is then added incoherently to the cross section \cite{Hernandez:PionNucleus,HybridRPWIA}.
We do not include this reaction here, hence we consider the results with (without) OSMM of the delta as a lower (upper) limit, so that the hybrid-RPWIA model is illustrated by the red band.
In principle, the decay width of the other resonances is also modified in the nuclear medium.
Including this effect is in the best of cases not free of ambiguities, because the other resonances are not as well known as the delta.
Anyhow, their contribution to the overall cross section is small, and approximately limited to the region $0.5~\mathrm{GeV} < P_{\pi} <1~\mathrm{GeV}$.
Therefore, the medium modification of the higher lying resonances is not taken into account, and can be considered as a (relatively) small uncertainty in our predictions.

It is interesting to compare the T2K-flux \cite{T2K:flux2016} with the neutrino fluxes in MiniBooNE and MINERvA, and confront the datasets in their comparison to the \mbox{hybrid-RPWIA} model predictions.
The T2K flux has a peak for neutrino energies around $600~\mathrm{MeV}$, comparable to the energy regime spanned by MiniBooNE \cite{MBflux:2009}.
However, the T2K-flux has a more significant high-energy tail.
This, along with the restrictions on lepton and pion kinematics in T2K, leads to the T2K data having a larger contribution from high energy neutrinos than the MiniBooNE data \cite{MB:pion}.
The MINERvA experiment spans a far larger energy range, the flux peaks around $3~\mathrm{GeV}$, and extends to about $20~\mathrm{GeV}$ \cite{NUMIflux}. 
Contrary to the MINERvA samples \cite{MINERvA:CCPI0,MINERvA:CC1PI}, there is no restriction on (reconstructed) quantities such as $W$ in the T2K data.
The influence of higher $W$ regions on the T2K dataset is demonstrated by comparing the results of the hybrid-RPWIA model to those of the LEM.
The cross section in the LEM for T2K kinematics is illustrated with the dash-dotted blue line in Fig.~\ref{fig:CS_Ppi}. 
The LEM with inclusion of Gaussian-dipole form factors for the resonances is also shown, labeled as ``LEM w/ FF''.
Both results are computed without OSMM.
In our previous work \cite{HybridRPWIA}, we found that for both the MiniBooNE and MINERvA CC1$\pi^+$ data, the hybrid-RPWIA and LEM (with or without form factors) did not differ significantly.
As mentioned, in the MINERvA data a cut is made restricting the phase space to $W < 1.4~\mathrm{GeV}$, thereby ensuring that the dominant reaction mechanism is delta-mediated pion production.
Due to the smooth transition between the LEM and the Regge approach, the hybrid-RPWIA model is identical to the LEM with form factors for $W\lesssim 1.5~\mathrm{GeV}$.
For the kinematics presented here however, we see important deviations between the different curves in Fig. \ref{fig:CS_Ppi}, showing that regions of higher $W$ contribute significantly to the T2K signal.
The cutoff form factors to regularize the resonant diagrams cure some of the pathological behavior, but a significant difference between the LEM with form factors and hybrid-RPWIA model still exists for $P_{\pi}>0.5~\mathrm{GeV}$.

In Figs.~\ref{fig:CS_Ppi}(b) and \ref{fig:CS_Ppi}(c), we see that the hybrid-RPWIA approach predicts a noteable smaller cross section over the whole kinematic range compared to the LEM.
When comparing the models in terms of $\cos\theta_{\mu}$ in Fig. \ref{fig:CS_Ppi}(d), we see that the main differences are found at forward scattering angles, consistent with regions of high $P_{\pi}$.

\subsection{\label{sec:results2}NuWro and Final State Interactions}
\begin{figure*}
\centering
\begin{large}
\resizebox{0.49\textwidth}{!}{\input{./gfx/CS_NuWro_Ppi_extended.tex}}
\resizebox{0.49\textwidth}{!}{\input{./gfx/CS_NuWro_THpi.tex}}
\end{large}
\caption{Single differential cross sections in terms of pion momentum (a) and scattering angle (b) compared to the CC1$\pi^+$ data reported by the T2K experiment \cite{T2KCC1PIH2O}.
The hybrid-RPWIA model is shown with a red band, where the lower and upper limits correspond to calculations with and without medium modification of the delta width respectively.
The NuWro cross section corresponding to the $1\pi1N$ final state, and the full NuWro calculation before and after FSI, both corresponding to the definition of the CC1$\pi^+$ signal, are shown.
We also show separately the contribution of coherent scattering in NuWro.}
\label{fig:CS_NuWro_Ppi}
\end{figure*}
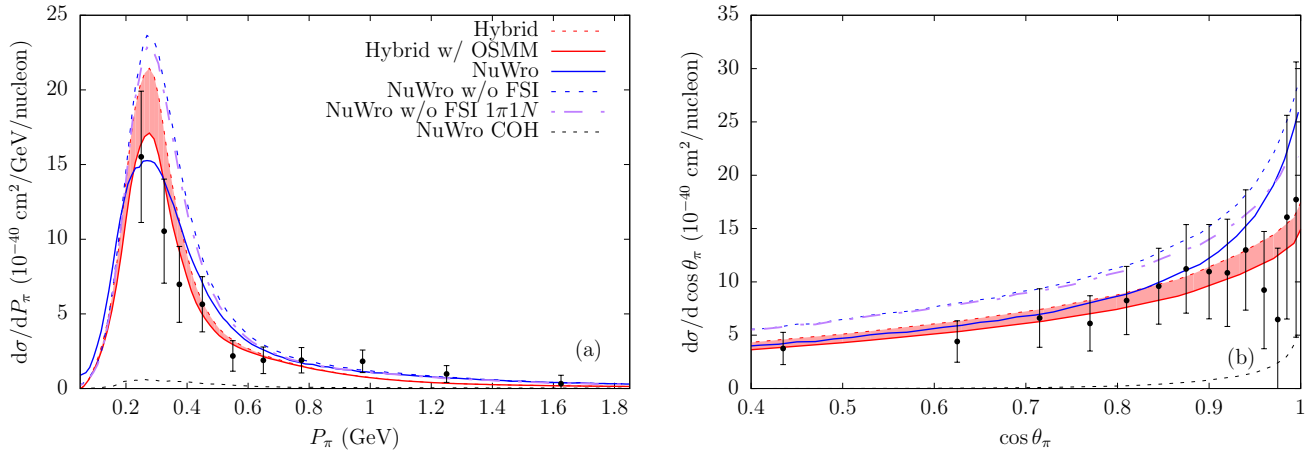

The hybrid-RPWIA model compares favorably to the T2K CC1$\pi^+$ data sample.
The total cross section reported by T2K is $\sigma_{\mathrm{tot}} = 4.25 \pm 0.48 (\mathrm{stat}) \pm 1.56 (\mathrm{syst})\times10^{-40} \text{cm}^2/\textrm{nucleon}$ \cite{T2KCC1PIH2O},
compatible with $\sigma_{\mathrm{tot}}=4.82\times10^{-40} \text{cm}^2/\textrm{nucleon}$ obtained with the hybrid-RPWIA model.
This result is the average of the predictions with and without OSMM, the uncertainty (as illustrated by the red band in Fig.~\ref{fig:CS_NuWro_Ppi}) due to medium modification of the delta width is around 9~\%.
However, these results do not include any FSI, which are expected to reduce the cross section due to absorption and charge-exchange of the produced pion.
Indeed, the NuWro Monte Carlo generator predicts a total cross section of $6.97 \times10^{-40} \text{cm}^2/\textrm{nucleon}$ before FSI, and $5.44 \times10^{-40} \text{cm}^2/\textrm{nucleon}$ after taking into account FSI.
In this section we judge the impact of FSI on the single differential cross sections in terms of muon and pion kinematics by comparing our results to NuWro calculations.
One should however be careful in estimating the effect of FSI by directly comparing both models because there are significant differences between them.

We use NuWro version 17.09, with default values for all of the parameters~\cite{NuWroSITE}.
The elementary SPP mechanism in NuWro, i.e. before FSI, consists of the delta resonance treated in the Adler-Rarita Schwinger model \cite{Adler}, parametrized by dipole form factors fitted to SPP data \cite{NuWroFF}.
A phenomenological non-resonant background is obtained from DIS, it is added incoherently to the resonant cross section \cite{NuWroDIS}. 
For $W > 1.6~\mathrm{GeV}$ a model based on DIS \cite{BODEKYANG,NuWroDIS} and Pythia hadronization routines is used~\cite{PYTHIA6}. 
A smooth transition from the resonance region to DIS is implemented for $W$ between $1.4$ and $1.6~\mathrm{GeV}$ \cite{WRONG}.

In NuWro events originating from quasi-elastic scattering (QE), meson-exchange currents (MEC), and coherent scattering are generated, in addition to the elementary SPP process.
The final state particles from these interactions, excluding those from coherent pion production, are propagated through the nuclear medium where they can undergo secondary interactions \cite{NuWroFSI}.

\begin{figure*}
\centering
\begin{large}
\resizebox{0.49\textwidth}{!}{\input{./gfx/CS_NuWro_Pmu_extended.tex}}
\resizebox{0.49\textwidth}{!}{\input{./gfx/CS_NuWro_THl.tex}}
\end{large}
\caption{Single differential cross sections in terms of $P_{\mu}$ (a), and $\cos\theta_{\mu}$ (b), compared to T2K CC1$\pi^+$ data \cite{T2KCC1PIH2O}. The labels are the same as in Fig.~\ref{fig:CS_NuWro_Ppi}.}
\label{fig:CS_NuWro_Pl}
\end{figure*}
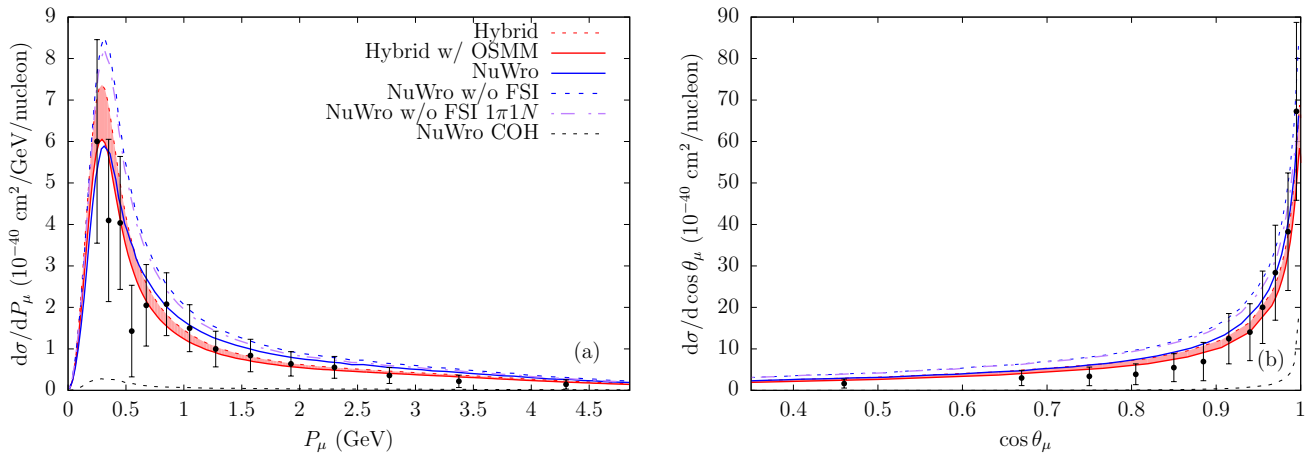

We compare the hybrid-RPWIA model to three results corresponding to different selection cuts in the NuWro simulation. 
First, we present the result where only a $\pi^+$ and a single nucleon are found in the hadronic final state,
before taking into account FSI.
This result, labeled as ``NuWro w/o FSI $1\pi1N$'' and depicted with the dash-dotted blue lines in Fig. \ref{fig:CS_NuWro_Ppi}, corresponds to the elementary SPP cross section described above, which should be comparable to the hybrid-RPWIA model.
The second result is labeled as ``NuWro w/o FSI'' and corresponds to the full calculation before FSI, where the hadronic final state is defined as a single $\pi^+$ and any number of nucleons (dashed blue lines in Fig. \ref{fig:CS_NuWro_Ppi}).
In practice, the main difference between the ``NuWro w/o FSI'', and the ``NuWro w/o FSI $1\pi1N$'' cross sections stems from the contribution of coherent scattering, which makes up around three percent of the former, and does not contribute to the latter.
We show the contribution of coherent scattering in NuWro separately, it is depicted with the dotted line and labeled as ``NuWro COH''.
Finally, the cross sections corresponding to the experimental signal after FSI are also shown.
The contributions of QE and MEC to the cross section are negligible, and the most important effect of FSI is a decrease of the cross section.
This final NuWro results corresponds to the solid blue line in Figs.~\ref{fig:CS_NuWro_Ppi} and \ref{fig:CS_NuWro_Pl}.

The influence of FSI on the $P_{\pi}$ distribution mainly consists of a strong reduction of the amount of pions with low momenta \cite{Hernandez:PionNucleus,Mosel:MB,HybridRPWIA}, this is shown in Fig. \ref{fig:CS_NuWro_Ppi}(a).
The NuWro $1\pi1N$ cross section is basically the same as the full cross section without FSI, the latter is slightly larger mainly due to the inclusion of coherent scattering.
In any case, the NuWro $1\pi1N$ cross-section is larger than the hybrid-RPWIA over the whole kinematic range.
This is consistent with our comparison to NuWro $1\pi^+$ calculations at MiniBooNE and MINERvA kinematics \cite{HybridRPWIA}.
The difference could be attributed to the form factors used to describe the couplings.
It was shown in Ref.~\cite{Gonzalez:SPPnucleon}, 
where both models are compared to SPP neutrino-deuterium data \cite{ANLBNLWilkinson},
that NuWro systematically obtains a larger total cross section.

The hybrid-RPWIA model tends to overestimate the number of pions at the lowest momenta, leaving room for FSI. 
However, a reduction of the low momentum peaks, as estimated by comparing to NuWro, would lead to the hybrid-RPWIA underpredicting the lowest momentum bins for both the $P_{\pi}$, and $P_{\mu}$ cross sections shown in Figs.~\ref{fig:CS_NuWro_Ppi}(a) and \ref{fig:CS_NuWro_Pl}(a).
This is also found in the GiBUU prediction of Ref.~\cite{Mosel:T2KH2O}, where it is argued that coherent pion production could provide additional strength in this region. 
 
In the comparison of the hybrid-RPWIA model with MiniBooNE and MINERvA $1\pi^+$ data, pion momenta up to approximately $500~\mathrm{MeV}$ were studied \cite{HybridRPWIA}.
Here, we see that the cross section in the high-$P_{\pi}$ region in NuWro, which is dominated by DIS, is larger than the Regge description in the hybrid-RPWIA model. 
The small cross section in the higher $P_{\pi}$ regions, relative to the data, may point to a lack of higher mass resonances \cite{Leitner:Resonances},
 or of high energy mechanisms that may contribute to the signal after FSI. 
The problem is tied to the description of the transition region.
Adding additional higher mass resonances would require unitarization of the amplitude in the LEM, thereby extending the validity of the LEM such that the transition region can be moved to larger values of $W$.

We show the comparison with the $\cos\theta_{\pi}$ distribution in Fig. \ref{fig:CS_NuWro_Ppi}(b).
The cross section in the hybrid-RPWIA model does not show the sharp rise at forward scattering angles present in the NuWro calculations.
It is in this kinematic region that contributions from coherent scattering and DIS are most important.
The effect of FSI is a constant reduction over the whole range of $\cos\theta_{\pi}$, except for the most forward angles where the reduction is not as strong as for the rest of the angular range.
This can be partly attributed to the fact that the coherent scattering events are not subject to FSI through the cascade in NuWro.
 
In Fig.~\ref{fig:CS_NuWro_Pl} the comparison with the data in terms of muon kinematics is shown.
Again, a difference in the overall strength in the cross section compared to NuWro is evident.
The high-$P_{\mu}$ tail is described well by the hybrid-RPWIA model, as is the low momentum peak.
As mentioned, one could expect a slight underestimation of the low-$P_{\mu}$ peak if FSI, as predicted by NuWro, would be included.
The same holds for the forward scattering cross section, in agreement with Ref. \cite{Mosel:T2KH2O}.
The region $\cos\theta_{\mu} < 0.9$ would however be in agreement with the data even after a reduction of the cross section due to FSI as estimated from the NuWro result.

\section{\label{sec:conclusion}Conclusions}
We compared the hybrid-RPWIA to the low energy model (LEM) and the T2K CC1$\pi^+$ data.
It is shown that a high energy model is necessary at T2K kinematics.
The contributions from the high energy tail of the flux are significant, and using the LEM leads to a sizeable overestimation of the cross section.
Introducing Gaussian-dipole form factors to regularize the resonant amplitudes cures some of the pathological behavior due to the resonant amplitudes far away from $s(u) \approx M_{res}^2$.
Still, the LEM with form factors overestimates the cross section for high pion momenta when compared to the hybrid-RPWIA model. 
This was not the case for the predictions at MiniBooNE and MINERvA CC1$\pi^+$ kinematics presented in Ref.~\cite{HybridRPWIA}, due to the cut on $W$ in MINERvA, and the smaller high energy contributions in MiniBooNE.

The shape of the single differential cross sections obtained within the hybrid-RPWIA model presented here are similar to the NuWro results, with the main exception being the forward pion scattering region.
It is in this region that the coherent and DIS contributions in NuWro predict a sharp rise.
When considering the size, we see that the hybrid-RPWIA model systematically predicts a lower cross section than the NuWro Monte Carlo generator.
These results are consistent with the previous comparisons shown in Refs.~\cite{Gonzalez:SPPnucleon,HybridRPWIA}. 

The general comparison of the hybrid-RPWIA model to data is favorable, the model reproduces the shape and strength of the data well, meanwhile leaving room for FSI at low pion and lepton momenta.
This is true for both the MINERvA, and T2K CC1$\pi^+$ data, but not for MiniBooNE, where the model clearly underpredicts the data \cite{HybridRPWIA}. 
There seems to be no obvious reason explaining why the hybrid-RPWIA model underestimates the MiniBooNE data for low values of $T_{\pi}$ (see Fig.~$5$ in Ref.~\cite{HybridRPWIA}),  while overpredicting T2K data in the same kinematic region, Fig.~\ref{fig:CS_NuWro_Ppi}(a).
Moreover, the NuWro predictions reproduce the MINERvA and T2K $\pi^+$ data nicely, while underestimating MiniBooNE.




\begin{acknowledgments}
This  work  was  supported  by  the  Interuniversity   Attraction   Poles   Programme   initiated   by
the  Belgian  Science  Policy  Office  (BriX  network
P7/12)  and  the  Research  Foundation  Flanders
(FWO-Flanders), and partially by the Special Research  Fund,  Ghent  University.
The  computational resources (Stevin Supercomputer Infrastructure)  and  services  used  in  this  work  were  provided  by  Ghent  University,  the  Hercules  Foundation   and   the   Flemish   Government.
K.N. and J.S.
were  partially  supported  by  the  Polish  National
Science  Center  (NCN),  under  Opus  Grant  No.
2016/21/B/ST2/01092, 
as well as by the Polish Ministry of Science and Higher Education, Grant No. DIR/WK/2017/05.
\end{acknowledgments}
\bibliographystyle{apsrev4-1.bst}
\bibliography{Bibliography}

\end{document}

%% file: gfx/CS_Ppi.tex
\begingroup
  \makeatletter
  \providecommand\color[2][]{%
    \GenericError{(gnuplot) \space\space\space\@spaces}{%
      Package color not loaded in conjunction with
      terminal option `colourtext'%
    }{See the gnuplot documentation for explanation.%
    }{Either use 'blacktext' in gnuplot or load the package
      color.sty in LaTeX.}%
    \renewcommand\color[2][]{}%
  }%
  \providecommand\includegraphics[2][]{%
    \GenericError{(gnuplot) \space\space\space\@spaces}{%
      Package graphicx or graphics not loaded%
    }{See the gnuplot documentation for explanation.%
    }{The gnuplot epslatex terminal needs graphicx.sty or graphics.sty.}%
    \renewcommand\includegraphics[2][]{}%
  }%
  \providecommand\rotatebox[2]{#2}%
  \@ifundefined{ifGPcolor}{%
    \newif\ifGPcolor
    \GPcolorfalse
  }{}%
  \@ifundefined{ifGPblacktext}{%
    \newif\ifGPblacktext
    \GPblacktexttrue
  }{}%
  \let\gplgaddtomacro\g@addto@macro
  \gdef\gplbacktext{}%
  \gdef\gplfronttext{}%
  \makeatother
  \ifGPblacktext
    \def\colorrgb#1{}%
    \def\colorgray#1{}%
  \else
    \ifGPcolor
      \def\colorrgb#1{\color[rgb]{#1}}%
      \def\colorgray#1{\color[gray]{#1}}%
      \expandafter\def\csname LTw\endcsname{\color{white}}%
      \expandafter\def\csname LTb\endcsname{\color{black}}%
      \expandafter\def\csname LTa\endcsname{\color{black}}%
      \expandafter\def\csname LT0\endcsname{\color[rgb]{1,0,0}}%
      \expandafter\def\csname LT1\endcsname{\color[rgb]{0,1,0}}%
      \expandafter\def\csname LT2\endcsname{\color[rgb]{0,0,1}}%
      \expandafter\def\csname LT3\endcsname{\color[rgb]{1,0,1}}%
      \expandafter\def\csname LT4\endcsname{\color[rgb]{0,1,1}}%
      \expandafter\def\csname LT5\endcsname{\color[rgb]{1,1,0}}%
      \expandafter\def\csname LT6\endcsname{\color[rgb]{0,0,0}}%
      \expandafter\def\csname LT7\endcsname{\color[rgb]{1,0.3,0}}%
      \expandafter\def\csname LT8\endcsname{\color[rgb]{0.5,0.5,0.5}}%
    \else
      \def\colorrgb#1{\color{black}}%
      \def\colorgray#1{\color[gray]{#1}}%
      \expandafter\def\csname LTw\endcsname{\color{white}}%
      \expandafter\def\csname LTb\endcsname{\color{black}}%
      \expandafter\def\csname LTa\endcsname{\color{black}}%
      \expandafter\def\csname LT0\endcsname{\color{black}}%
      \expandafter\def\csname LT1\endcsname{\color{black}}%
      \expandafter\def\csname LT2\endcsname{\color{black}}%
      \expandafter\def\csname LT3\endcsname{\color{black}}%
      \expandafter\def\csname LT4\endcsname{\color{black}}%
      \expandafter\def\csname LT5\endcsname{\color{black}}%
      \expandafter\def\csname LT6\endcsname{\color{black}}%
      \expandafter\def\csname LT7\endcsname{\color{black}}%
      \expandafter\def\csname LT8\endcsname{\color{black}}%
    \fi
  \fi
    \setlength{\unitlength}{0.0500bp}%
    \ifx\gptboxheight\undefined%
      \newlength{\gptboxheight}%
      \newlength{\gptboxwidth}%
      \newsavebox{\gptboxtext}%
    \fi%
    \setlength{\fboxrule}{0.5pt}%
    \setlength{\fboxsep}{1pt}%
\begin{picture}(7200.00,5040.00)%
    \gplgaddtomacro\gplbacktext{%
      \csname LTb\endcsname%
      \put(682,704){\makebox(0,0)[r]{\strut{}$0$}}%
      \put(682,1518){\makebox(0,0)[r]{\strut{}$5$}}%
      \put(682,2332){\makebox(0,0)[r]{\strut{}$10$}}%
      \put(682,3147){\makebox(0,0)[r]{\strut{}$15$}}%
      \put(682,3961){\makebox(0,0)[r]{\strut{}$20$}}%
      \put(682,4775){\makebox(0,0)[r]{\strut{}$25$}}%
      \put(1156,484){\makebox(0,0){\strut{}$0.2$}}%
      \put(1841,484){\makebox(0,0){\strut{}$0.4$}}%
      \put(2525,484){\makebox(0,0){\strut{}$0.6$}}%
      \put(3210,484){\makebox(0,0){\strut{}$0.8$}}%
      \put(3894,484){\makebox(0,0){\strut{}$1$}}%
      \put(4579,484){\makebox(0,0){\strut{}$1.2$}}%
      \put(5263,484){\makebox(0,0){\strut{}$1.4$}}%
      \put(5947,484){\makebox(0,0){\strut{}$1.6$}}%
      \put(6632,484){\makebox(0,0){\strut{}$1.8$}}%
      \put(6204,1111){\makebox(0,0)[l]{\strut{}(a)}}%
    }%
    \gplgaddtomacro\gplfronttext{%
      \csname LTb\endcsname%
      \put(176,2739){\rotatebox{-270}{\makebox(0,0){\strut{}$\mathrm{d}\sigma/\mathrm{d}P_{\pi}$  ($10^{-40} \mathrm{cm}^2/\mathrm{GeV}/\mathrm{nucleon}$)}}}%
      \put(3808,154){\makebox(0,0){\strut{}$P_{\pi}$  (GeV)}}%
      \csname LTb\endcsname%
      \put(5816,4602){\makebox(0,0)[r]{\strut{}Hybrid}}%
      \csname LTb\endcsname%
      \put(5816,4382){\makebox(0,0)[r]{\strut{}Hybrid w/ MM}}%
      \csname LTb\endcsname%
      \put(5816,4162){\makebox(0,0)[r]{\strut{}LEM w/ FF}}%
      \csname LTb\endcsname%
      \put(5816,3942){\makebox(0,0)[r]{\strut{}LEM}}%
      \csname LTb\endcsname%
      \put(5816,3722){\makebox(0,0)[r]{\strut{}Delta+BG}}%
    }%
    \gplbacktext
    \put(0,0){\includegraphics{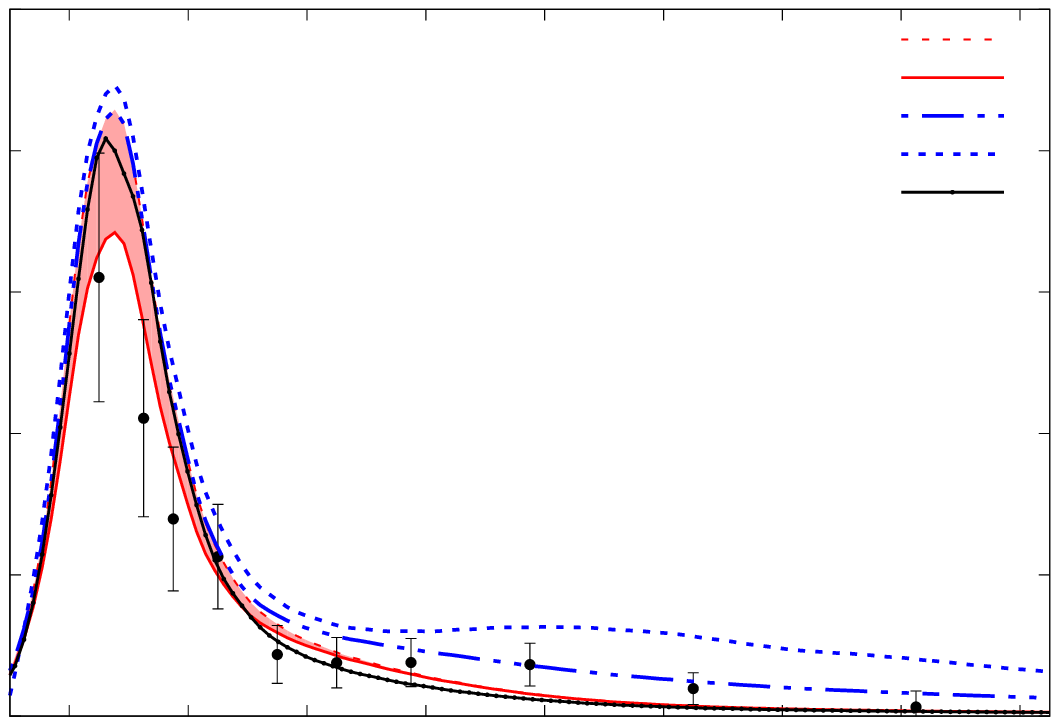}}%
    \gplfronttext
  \end{picture}%
\endgroup

%% file: gfx/CS_Pmu.tex
\begingroup
  \makeatletter
  \providecommand\color[2][]{%
    \GenericError{(gnuplot) \space\space\space\@spaces}{%
      Package color not loaded in conjunction with
      terminal option `colourtext'%
    }{See the gnuplot documentation for explanation.%
    }{Either use 'blacktext' in gnuplot or load the package
      color.sty in LaTeX.}%
    \renewcommand\color[2][]{}%
  }%
  \providecommand\includegraphics[2][]{%
    \GenericError{(gnuplot) \space\space\space\@spaces}{%
      Package graphicx or graphics not loaded%
    }{See the gnuplot documentation for explanation.%
    }{The gnuplot epslatex terminal needs graphicx.sty or graphics.sty.}%
    \renewcommand\includegraphics[2][]{}%
  }%
  \providecommand\rotatebox[2]{#2}%
  \@ifundefined{ifGPcolor}{%
    \newif\ifGPcolor
    \GPcolorfalse
  }{}%
  \@ifundefined{ifGPblacktext}{%
    \newif\ifGPblacktext
    \GPblacktexttrue
  }{}%
  \let\gplgaddtomacro\g@addto@macro
  \gdef\gplbacktext{}%
  \gdef\gplfronttext{}%
  \makeatother
  \ifGPblacktext
    \def\colorrgb#1{}%
    \def\colorgray#1{}%
  \else
    \ifGPcolor
      \def\colorrgb#1{\color[rgb]{#1}}%
      \def\colorgray#1{\color[gray]{#1}}%
      \expandafter\def\csname LTw\endcsname{\color{white}}%
      \expandafter\def\csname LTb\endcsname{\color{black}}%
      \expandafter\def\csname LTa\endcsname{\color{black}}%
      \expandafter\def\csname LT0\endcsname{\color[rgb]{1,0,0}}%
      \expandafter\def\csname LT1\endcsname{\color[rgb]{0,1,0}}%
      \expandafter\def\csname LT2\endcsname{\color[rgb]{0,0,1}}%
      \expandafter\def\csname LT3\endcsname{\color[rgb]{1,0,1}}%
      \expandafter\def\csname LT4\endcsname{\color[rgb]{0,1,1}}%
      \expandafter\def\csname LT5\endcsname{\color[rgb]{1,1,0}}%
      \expandafter\def\csname LT6\endcsname{\color[rgb]{0,0,0}}%
      \expandafter\def\csname LT7\endcsname{\color[rgb]{1,0.3,0}}%
      \expandafter\def\csname LT8\endcsname{\color[rgb]{0.5,0.5,0.5}}%
    \else
      \def\colorrgb#1{\color{black}}%
      \def\colorgray#1{\color[gray]{#1}}%
      \expandafter\def\csname LTw\endcsname{\color{white}}%
      \expandafter\def\csname LTb\endcsname{\color{black}}%
      \expandafter\def\csname LTa\endcsname{\color{black}}%
      \expandafter\def\csname LT0\endcsname{\color{black}}%
      \expandafter\def\csname LT1\endcsname{\color{black}}%
      \expandafter\def\csname LT2\endcsname{\color{black}}%
      \expandafter\def\csname LT3\endcsname{\color{black}}%
      \expandafter\def\csname LT4\endcsname{\color{black}}%
      \expandafter\def\csname LT5\endcsname{\color{black}}%
      \expandafter\def\csname LT6\endcsname{\color{black}}%
      \expandafter\def\csname LT7\endcsname{\color{black}}%
      \expandafter\def\csname LT8\endcsname{\color{black}}%
    \fi
  \fi
    \setlength{\unitlength}{0.0500bp}%
    \ifx\gptboxheight\undefined%
      \newlength{\gptboxheight}%
      \newlength{\gptboxwidth}%
      \newsavebox{\gptboxtext}%
    \fi%
    \setlength{\fboxrule}{0.5pt}%
    \setlength{\fboxsep}{1pt}%
\begin{picture}(7200.00,5040.00)%
    \gplgaddtomacro\gplbacktext{%
      \csname LTb\endcsname%
      \put(682,704){\makebox(0,0)[r]{\strut{}$0$}}%
      \put(682,1111){\makebox(0,0)[r]{\strut{}$1$}}%
      \put(682,1518){\makebox(0,0)[r]{\strut{}$2$}}%
      \put(682,1925){\makebox(0,0)[r]{\strut{}$3$}}%
      \put(682,2332){\makebox(0,0)[r]{\strut{}$4$}}%
      \put(682,2740){\makebox(0,0)[r]{\strut{}$5$}}%
      \put(682,3147){\makebox(0,0)[r]{\strut{}$6$}}%
      \put(682,3554){\makebox(0,0)[r]{\strut{}$7$}}%
      \put(682,3961){\makebox(0,0)[r]{\strut{}$8$}}%
      \put(682,4368){\makebox(0,0)[r]{\strut{}$9$}}%
      \put(682,4775){\makebox(0,0)[r]{\strut{}$10$}}%
      \put(814,484){\makebox(0,0){\strut{}$0$}}%
      \put(1479,484){\makebox(0,0){\strut{}$0.5$}}%
      \put(2145,484){\makebox(0,0){\strut{}$1$}}%
      \put(2810,484){\makebox(0,0){\strut{}$1.5$}}%
      \put(3476,484){\makebox(0,0){\strut{}$2$}}%
      \put(4141,484){\makebox(0,0){\strut{}$2.5$}}%
      \put(4807,484){\makebox(0,0){\strut{}$3$}}%
      \put(5472,484){\makebox(0,0){\strut{}$3.5$}}%
      \put(6138,484){\makebox(0,0){\strut{}$4$}}%
      \put(6803,484){\makebox(0,0){\strut{}$4.5$}}%
      \put(6204,1111){\makebox(0,0)[l]{\strut{}(b)}}%
    }%
    \gplgaddtomacro\gplfronttext{%
      \csname LTb\endcsname%
      \put(176,2739){\rotatebox{-270}{\makebox(0,0){\strut{}$\mathrm{d}\sigma/\mathrm{d}P_{\mu}$  ($10^{-40}\mathrm{cm}^2/\mathrm{GeV}/\mathrm{nucleon}$)}}}%
      \put(3808,154){\makebox(0,0){\strut{}$P_{\mu}$  (GeV)}}%
    }%
    \gplbacktext
    \put(0,0){\includegraphics{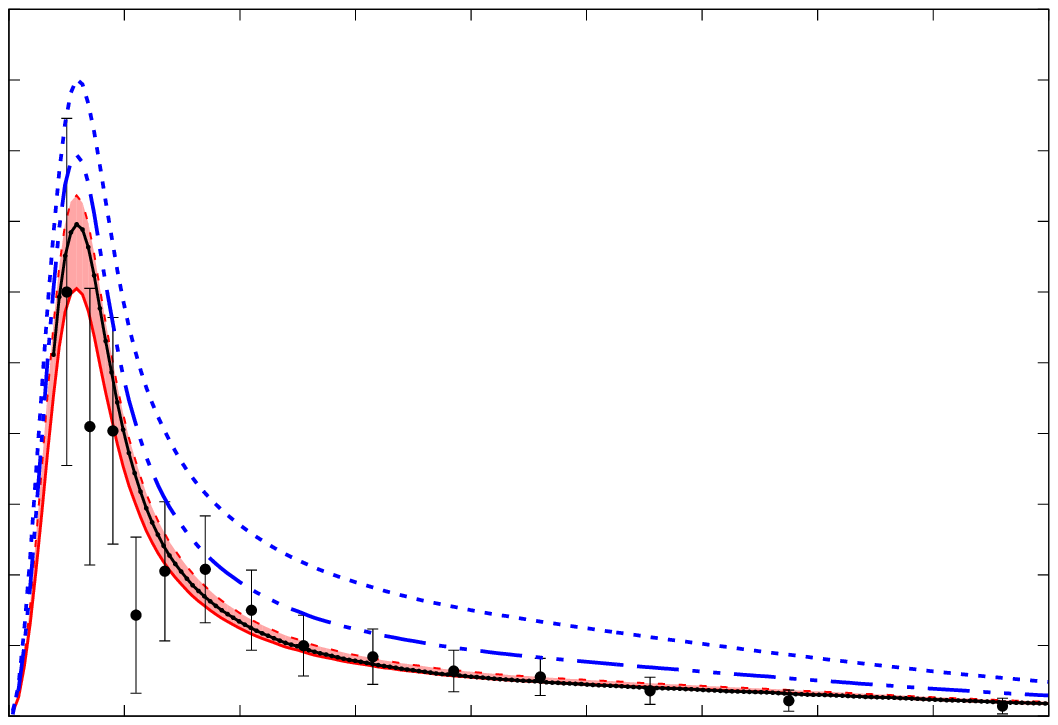}}%
    \gplfronttext
  \end{picture}%
\endgroup

%% file: gfx/CS_THpi.tex
\begingroup
  \makeatletter
  \providecommand\color[2][]{%
    \GenericError{(gnuplot) \space\space\space\@spaces}{%
      Package color not loaded in conjunction with
      terminal option `colourtext'%
    }{See the gnuplot documentation for explanation.%
    }{Either use 'blacktext' in gnuplot or load the package
      color.sty in LaTeX.}%
    \renewcommand\color[2][]{}%
  }%
  \providecommand\includegraphics[2][]{%
    \GenericError{(gnuplot) \space\space\space\@spaces}{%
      Package graphicx or graphics not loaded%
    }{See the gnuplot documentation for explanation.%
    }{The gnuplot epslatex terminal needs graphicx.sty or graphics.sty.}%
    \renewcommand\includegraphics[2][]{}%
  }%
  \providecommand\rotatebox[2]{#2}%
  \@ifundefined{ifGPcolor}{%
    \newif\ifGPcolor
    \GPcolorfalse
  }{}%
  \@ifundefined{ifGPblacktext}{%
    \newif\ifGPblacktext
    \GPblacktexttrue
  }{}%
  \let\gplgaddtomacro\g@addto@macro
  \gdef\gplbacktext{}%
  \gdef\gplfronttext{}%
  \makeatother
  \ifGPblacktext
    \def\colorrgb#1{}%
    \def\colorgray#1{}%
  \else
    \ifGPcolor
      \def\colorrgb#1{\color[rgb]{#1}}%
      \def\colorgray#1{\color[gray]{#1}}%
      \expandafter\def\csname LTw\endcsname{\color{white}}%
      \expandafter\def\csname LTb\endcsname{\color{black}}%
      \expandafter\def\csname LTa\endcsname{\color{black}}%
      \expandafter\def\csname LT0\endcsname{\color[rgb]{1,0,0}}%
      \expandafter\def\csname LT1\endcsname{\color[rgb]{0,1,0}}%
      \expandafter\def\csname LT2\endcsname{\color[rgb]{0,0,1}}%
      \expandafter\def\csname LT3\endcsname{\color[rgb]{1,0,1}}%
      \expandafter\def\csname LT4\endcsname{\color[rgb]{0,1,1}}%
      \expandafter\def\csname LT5\endcsname{\color[rgb]{1,1,0}}%
      \expandafter\def\csname LT6\endcsname{\color[rgb]{0,0,0}}%
      \expandafter\def\csname LT7\endcsname{\color[rgb]{1,0.3,0}}%
      \expandafter\def\csname LT8\endcsname{\color[rgb]{0.5,0.5,0.5}}%
    \else
      \def\colorrgb#1{\color{black}}%
      \def\colorgray#1{\color[gray]{#1}}%
      \expandafter\def\csname LTw\endcsname{\color{white}}%
      \expandafter\def\csname LTb\endcsname{\color{black}}%
      \expandafter\def\csname LTa\endcsname{\color{black}}%
      \expandafter\def\csname LT0\endcsname{\color{black}}%
      \expandafter\def\csname LT1\endcsname{\color{black}}%
      \expandafter\def\csname LT2\endcsname{\color{black}}%
      \expandafter\def\csname LT3\endcsname{\color{black}}%
      \expandafter\def\csname LT4\endcsname{\color{black}}%
      \expandafter\def\csname LT5\endcsname{\color{black}}%
      \expandafter\def\csname LT6\endcsname{\color{black}}%
      \expandafter\def\csname LT7\endcsname{\color{black}}%
      \expandafter\def\csname LT8\endcsname{\color{black}}%
    \fi
  \fi
    \setlength{\unitlength}{0.0500bp}%
    \ifx\gptboxheight\undefined%
      \newlength{\gptboxheight}%
      \newlength{\gptboxwidth}%
      \newsavebox{\gptboxtext}%
    \fi%
    \setlength{\fboxrule}{0.5pt}%
    \setlength{\fboxsep}{1pt}%
\begin{picture}(7200.00,5040.00)%
    \gplgaddtomacro\gplbacktext{%
      \csname LTb\endcsname%
      \put(682,704){\makebox(0,0)[r]{\strut{}$0$}}%
      \put(682,1286){\makebox(0,0)[r]{\strut{}$5$}}%
      \put(682,1867){\makebox(0,0)[r]{\strut{}$10$}}%
      \put(682,2449){\makebox(0,0)[r]{\strut{}$15$}}%
      \put(682,3030){\makebox(0,0)[r]{\strut{}$20$}}%
      \put(682,3612){\makebox(0,0)[r]{\strut{}$25$}}%
      \put(682,4193){\makebox(0,0)[r]{\strut{}$30$}}%
      \put(682,4775){\makebox(0,0)[r]{\strut{}$35$}}%
      \put(814,484){\makebox(0,0){\strut{}$0.4$}}%
      \put(1812,484){\makebox(0,0){\strut{}$0.5$}}%
      \put(2810,484){\makebox(0,0){\strut{}$0.6$}}%
      \put(3808,484){\makebox(0,0){\strut{}$0.7$}}%
      \put(4807,484){\makebox(0,0){\strut{}$0.8$}}%
      \put(5805,484){\makebox(0,0){\strut{}$0.9$}}%
      \put(6803,484){\makebox(0,0){\strut{}$1$}}%
      \put(6000,1030){\makebox(0,0)[l]{\strut{}(c)}}%
    }%
    \gplgaddtomacro\gplfronttext{%
      \csname LTb\endcsname%
      \put(176,2739){\rotatebox{-270}{\makebox(0,0){\strut{}$\mathrm{d}\sigma/\mathrm{d}\cos\theta_{\pi}$  ($10^{-40}~\mathrm{cm}^{2}/\mathrm{nucleon}$)}}}%
      \put(3808,154){\makebox(0,0){\strut{}$\cos\theta_{\pi}$}}%
    }%
    \gplbacktext
    \put(0,0){\includegraphics{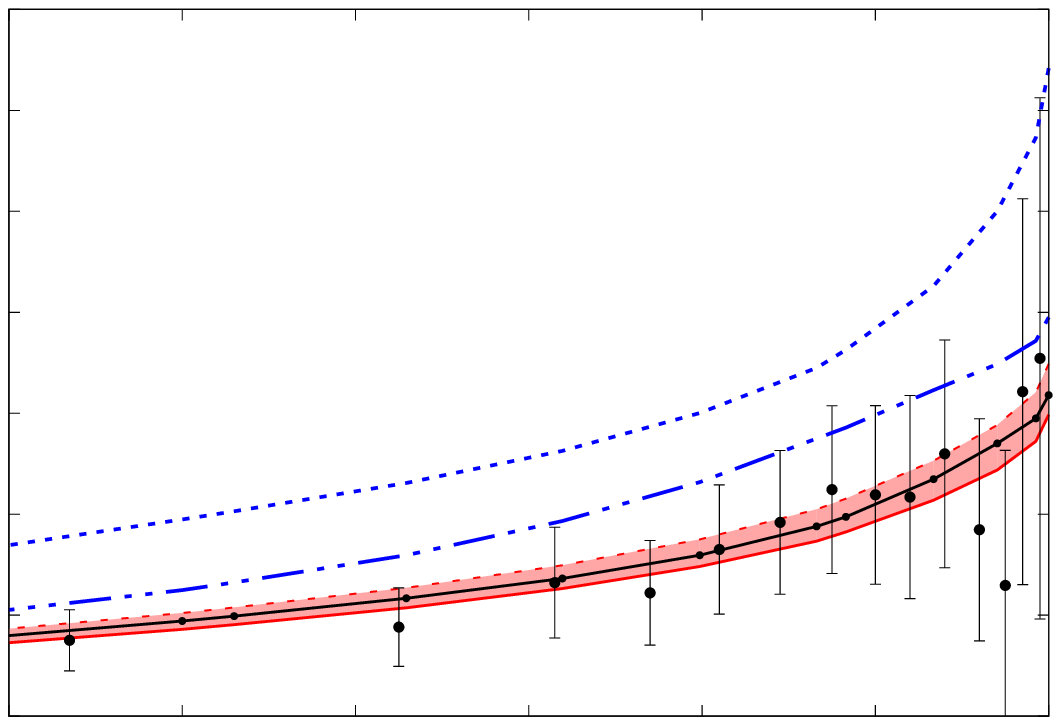}}%
    \gplfronttext
  \end{picture}%
\endgroup

%% file: gfx/CS_THl.tex
\begingroup
  \makeatletter
  \providecommand\color[2][]{%
    \GenericError{(gnuplot) \space\space\space\@spaces}{%
      Package color not loaded in conjunction with
      terminal option `colourtext'%
    }{See the gnuplot documentation for explanation.%
    }{Either use 'blacktext' in gnuplot or load the package
      color.sty in LaTeX.}%
    \renewcommand\color[2][]{}%
  }%
  \providecommand\includegraphics[2][]{%
    \GenericError{(gnuplot) \space\space\space\@spaces}{%
      Package graphicx or graphics not loaded%
    }{See the gnuplot documentation for explanation.%
    }{The gnuplot epslatex terminal needs graphicx.sty or graphics.sty.}%
    \renewcommand\includegraphics[2][]{}%
  }%
  \providecommand\rotatebox[2]{#2}%
  \@ifundefined{ifGPcolor}{%
    \newif\ifGPcolor
    \GPcolorfalse
  }{}%
  \@ifundefined{ifGPblacktext}{%
    \newif\ifGPblacktext
    \GPblacktexttrue
  }{}%
  \let\gplgaddtomacro\g@addto@macro
  \gdef\gplbacktext{}%
  \gdef\gplfronttext{}%
  \makeatother
  \ifGPblacktext
    \def\colorrgb#1{}%
    \def\colorgray#1{}%
  \else
    \ifGPcolor
      \def\colorrgb#1{\color[rgb]{#1}}%
      \def\colorgray#1{\color[gray]{#1}}%
      \expandafter\def\csname LTw\endcsname{\color{white}}%
      \expandafter\def\csname LTb\endcsname{\color{black}}%
      \expandafter\def\csname LTa\endcsname{\color{black}}%
      \expandafter\def\csname LT0\endcsname{\color[rgb]{1,0,0}}%
      \expandafter\def\csname LT1\endcsname{\color[rgb]{0,1,0}}%
      \expandafter\def\csname LT2\endcsname{\color[rgb]{0,0,1}}%
      \expandafter\def\csname LT3\endcsname{\color[rgb]{1,0,1}}%
      \expandafter\def\csname LT4\endcsname{\color[rgb]{0,1,1}}%
      \expandafter\def\csname LT5\endcsname{\color[rgb]{1,1,0}}%
      \expandafter\def\csname LT6\endcsname{\color[rgb]{0,0,0}}%
      \expandafter\def\csname LT7\endcsname{\color[rgb]{1,0.3,0}}%
      \expandafter\def\csname LT8\endcsname{\color[rgb]{0.5,0.5,0.5}}%
    \else
      \def\colorrgb#1{\color{black}}%
      \def\colorgray#1{\color[gray]{#1}}%
      \expandafter\def\csname LTw\endcsname{\color{white}}%
      \expandafter\def\csname LTb\endcsname{\color{black}}%
      \expandafter\def\csname LTa\endcsname{\color{black}}%
      \expandafter\def\csname LT0\endcsname{\color{black}}%
      \expandafter\def\csname LT1\endcsname{\color{black}}%
      \expandafter\def\csname LT2\endcsname{\color{black}}%
      \expandafter\def\csname LT3\endcsname{\color{black}}%
      \expandafter\def\csname LT4\endcsname{\color{black}}%
      \expandafter\def\csname LT5\endcsname{\color{black}}%
      \expandafter\def\csname LT6\endcsname{\color{black}}%
      \expandafter\def\csname LT7\endcsname{\color{black}}%
      \expandafter\def\csname LT8\endcsname{\color{black}}%
    \fi
  \fi
    \setlength{\unitlength}{0.0500bp}%
    \ifx\gptboxheight\undefined%
      \newlength{\gptboxheight}%
      \newlength{\gptboxwidth}%
      \newsavebox{\gptboxtext}%
    \fi%
    \setlength{\fboxrule}{0.5pt}%
    \setlength{\fboxsep}{1pt}%
\begin{picture}(7200.00,5040.00)%
    \gplgaddtomacro\gplbacktext{%
      \csname LTb\endcsname%
      \put(682,704){\makebox(0,0)[r]{\strut{}$0$}}%
      \put(682,1247){\makebox(0,0)[r]{\strut{}$10$}}%
      \put(682,1790){\makebox(0,0)[r]{\strut{}$20$}}%
      \put(682,2332){\makebox(0,0)[r]{\strut{}$30$}}%
      \put(682,2875){\makebox(0,0)[r]{\strut{}$40$}}%
      \put(682,3418){\makebox(0,0)[r]{\strut{}$50$}}%
      \put(682,3961){\makebox(0,0)[r]{\strut{}$60$}}%
      \put(682,4504){\makebox(0,0)[r]{\strut{}$70$}}%
      \put(814,484){\makebox(0,0){\strut{}$0.4$}}%
      \put(1812,484){\makebox(0,0){\strut{}$0.5$}}%
      \put(2810,484){\makebox(0,0){\strut{}$0.6$}}%
      \put(3808,484){\makebox(0,0){\strut{}$0.7$}}%
      \put(4807,484){\makebox(0,0){\strut{}$0.8$}}%
      \put(5805,484){\makebox(0,0){\strut{}$0.9$}}%
      \put(6803,484){\makebox(0,0){\strut{}$1$}}%
      \put(6324,1030){\makebox(0,0)[l]{\strut{}(d)}}%
    }%
    \gplgaddtomacro\gplfronttext{%
      \csname LTb\endcsname%
      \put(176,2739){\rotatebox{-270}{\makebox(0,0){\strut{}$\mathrm{d}\sigma/\mathrm{d}\cos\theta_{\mu}$  ($10^{-40}~\mathrm{cm}^{2}/\mathrm{nucleon}$)}}}%
      \put(3808,154){\makebox(0,0){\strut{}$\cos\theta_{\mu}$}}%
    }%
    \gplbacktext
    \put(0,0){\includegraphics{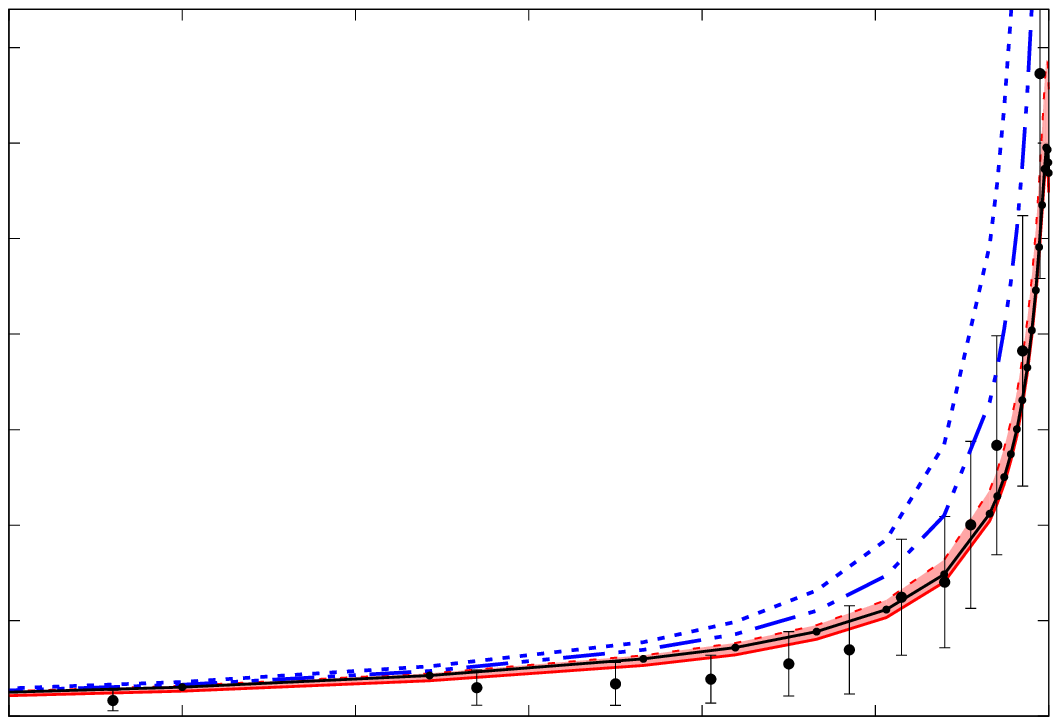}}%
    \gplfronttext
  \end{picture}%
\endgroup

%% file: gfx/CS_NuWro_Ppi_extended.tex
\begingroup
  \makeatletter
  \providecommand\color[2][]{%
    \GenericError{(gnuplot) \space\space\space\@spaces}{%
      Package color not loaded in conjunction with
      terminal option `colourtext'%
    }{See the gnuplot documentation for explanation.%
    }{Either use 'blacktext' in gnuplot or load the package
      color.sty in LaTeX.}%
    \renewcommand\color[2][]{}%
  }%
  \providecommand\includegraphics[2][]{%
    \GenericError{(gnuplot) \space\space\space\@spaces}{%
      Package graphicx or graphics not loaded%
    }{See the gnuplot documentation for explanation.%
    }{The gnuplot epslatex terminal needs graphicx.sty or graphics.sty.}%
    \renewcommand\includegraphics[2][]{}%
  }%
  \providecommand\rotatebox[2]{#2}%
  \@ifundefined{ifGPcolor}{%
    \newif\ifGPcolor
    \GPcolorfalse
  }{}%
  \@ifundefined{ifGPblacktext}{%
    \newif\ifGPblacktext
    \GPblacktexttrue
  }{}%
  \let\gplgaddtomacro\g@addto@macro
  \gdef\gplbacktext{}%
  \gdef\gplfronttext{}%
  \makeatother
  \ifGPblacktext
    \def\colorrgb#1{}%
    \def\colorgray#1{}%
  \else
    \ifGPcolor
      \def\colorrgb#1{\color[rgb]{#1}}%
      \def\colorgray#1{\color[gray]{#1}}%
      \expandafter\def\csname LTw\endcsname{\color{white}}%
      \expandafter\def\csname LTb\endcsname{\color{black}}%
      \expandafter\def\csname LTa\endcsname{\color{black}}%
      \expandafter\def\csname LT0\endcsname{\color[rgb]{1,0,0}}%
      \expandafter\def\csname LT1\endcsname{\color[rgb]{0,1,0}}%
      \expandafter\def\csname LT2\endcsname{\color[rgb]{0,0,1}}%
      \expandafter\def\csname LT3\endcsname{\color[rgb]{1,0,1}}%
      \expandafter\def\csname LT4\endcsname{\color[rgb]{0,1,1}}%
      \expandafter\def\csname LT5\endcsname{\color[rgb]{1,1,0}}%
      \expandafter\def\csname LT6\endcsname{\color[rgb]{0,0,0}}%
      \expandafter\def\csname LT7\endcsname{\color[rgb]{1,0.3,0}}%
      \expandafter\def\csname LT8\endcsname{\color[rgb]{0.5,0.5,0.5}}%
    \else
      \def\colorrgb#1{\color{black}}%
      \def\colorgray#1{\color[gray]{#1}}%
      \expandafter\def\csname LTw\endcsname{\color{white}}%
      \expandafter\def\csname LTb\endcsname{\color{black}}%
      \expandafter\def\csname LTa\endcsname{\color{black}}%
      \expandafter\def\csname LT0\endcsname{\color{black}}%
      \expandafter\def\csname LT1\endcsname{\color{black}}%
      \expandafter\def\csname LT2\endcsname{\color{black}}%
      \expandafter\def\csname LT3\endcsname{\color{black}}%
      \expandafter\def\csname LT4\endcsname{\color{black}}%
      \expandafter\def\csname LT5\endcsname{\color{black}}%
      \expandafter\def\csname LT6\endcsname{\color{black}}%
      \expandafter\def\csname LT7\endcsname{\color{black}}%
      \expandafter\def\csname LT8\endcsname{\color{black}}%
    \fi
  \fi
    \setlength{\unitlength}{0.0500bp}%
    \ifx\gptboxheight\undefined%
      \newlength{\gptboxheight}%
      \newlength{\gptboxwidth}%
      \newsavebox{\gptboxtext}%
    \fi%
    \setlength{\fboxrule}{0.5pt}%
    \setlength{\fboxsep}{1pt}%
\begin{picture}(7200.00,5040.00)%
    \gplgaddtomacro\gplbacktext{%
      \csname LTb\endcsname%
      \put(682,704){\makebox(0,0)[r]{\strut{}$0$}}%
      \put(682,1518){\makebox(0,0)[r]{\strut{}$5$}}%
      \put(682,2332){\makebox(0,0)[r]{\strut{}$10$}}%
      \put(682,3147){\makebox(0,0)[r]{\strut{}$15$}}%
      \put(682,3961){\makebox(0,0)[r]{\strut{}$20$}}%
      \put(682,4775){\makebox(0,0)[r]{\strut{}$25$}}%
      \put(1313,484){\makebox(0,0){\strut{}$0.2$}}%
      \put(1979,484){\makebox(0,0){\strut{}$0.4$}}%
      \put(2644,484){\makebox(0,0){\strut{}$0.6$}}%
      \put(3309,484){\makebox(0,0){\strut{}$0.8$}}%
      \put(3975,484){\makebox(0,0){\strut{}$1$}}%
      \put(4640,484){\makebox(0,0){\strut{}$1.2$}}%
      \put(5306,484){\makebox(0,0){\strut{}$1.4$}}%
      \put(5971,484){\makebox(0,0){\strut{}$1.6$}}%
      \put(6637,484){\makebox(0,0){\strut{}$1.8$}}%
      \put(6204,1111){\makebox(0,0)[l]{\strut{}(a)}}%
    }%
    \gplgaddtomacro\gplfronttext{%
      \csname LTb\endcsname%
      \put(176,2739){\rotatebox{-270}{\makebox(0,0){\strut{}$\mathrm{d}\sigma/\mathrm{d}P_{\pi}$  ($10^{-40}~\mathrm{cm}^{2}/\mathrm{GeV}/\mathrm{nucleon}$)}}}%
      \put(3808,154){\makebox(0,0){\strut{}$P_{\pi}$  (GeV)}}%
      \csname LTb\endcsname%
      \put(5816,4602){\makebox(0,0)[r]{\strut{}Hybrid}}%
      \csname LTb\endcsname%
      \put(5816,4382){\makebox(0,0)[r]{\strut{}Hybrid w/ OSMM}}%
      \csname LTb\endcsname%
      \put(5816,4162){\makebox(0,0)[r]{\strut{}NuWro}}%
      \csname LTb\endcsname%
      \put(5816,3942){\makebox(0,0)[r]{\strut{}NuWro w/o FSI}}%
      \csname LTb\endcsname%
      \put(5816,3722){\makebox(0,0)[r]{\strut{}NuWro w/o FSI $1\pi1N$}}%
      \csname LTb\endcsname%
      \put(5816,3502){\makebox(0,0)[r]{\strut{}NuWro COH}}%
    }%
    \gplbacktext
    \put(0,0){\includegraphics{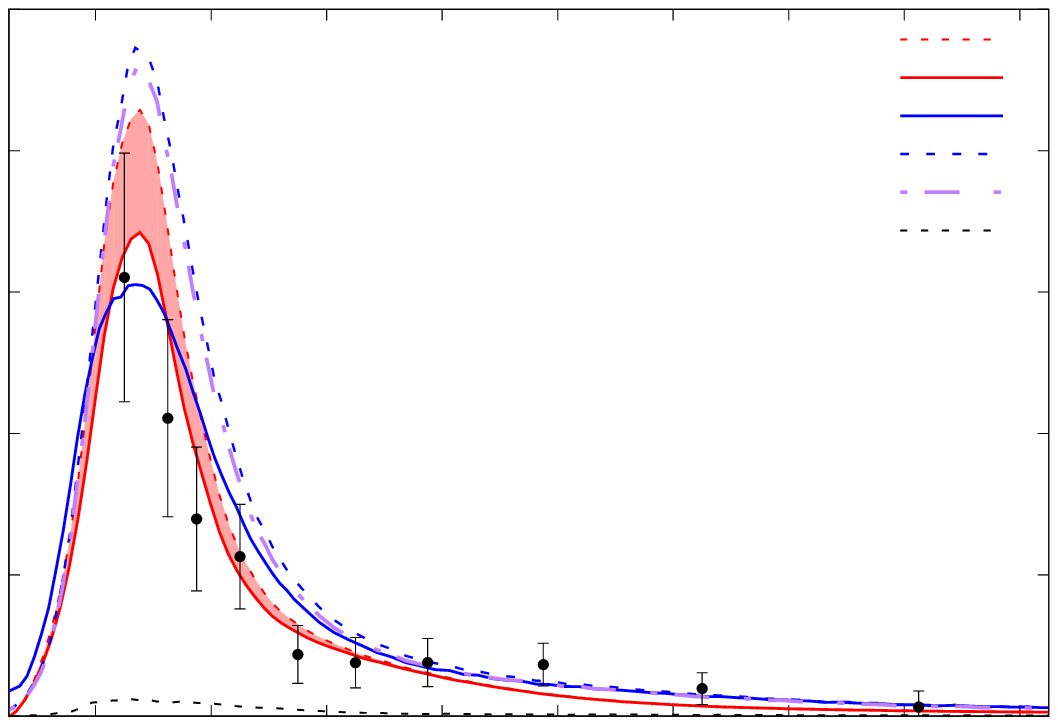}}%
    \gplfronttext
  \end{picture}%
\endgroup

%% file: gfx/CS_NuWro_THpi.tex
\begingroup
  \makeatletter
  \providecommand\color[2][]{%
    \GenericError{(gnuplot) \space\space\space\@spaces}{%
      Package color not loaded in conjunction with
      terminal option `colourtext'%
    }{See the gnuplot documentation for explanation.%
    }{Either use 'blacktext' in gnuplot or load the package
      color.sty in LaTeX.}%
    \renewcommand\color[2][]{}%
  }%
  \providecommand\includegraphics[2][]{%
    \GenericError{(gnuplot) \space\space\space\@spaces}{%
      Package graphicx or graphics not loaded%
    }{See the gnuplot documentation for explanation.%
    }{The gnuplot epslatex terminal needs graphicx.sty or graphics.sty.}%
    \renewcommand\includegraphics[2][]{}%
  }%
  \providecommand\rotatebox[2]{#2}%
  \@ifundefined{ifGPcolor}{%
    \newif\ifGPcolor
    \GPcolorfalse
  }{}%
  \@ifundefined{ifGPblacktext}{%
    \newif\ifGPblacktext
    \GPblacktexttrue
  }{}%
  \let\gplgaddtomacro\g@addto@macro
  \gdef\gplbacktext{}%
  \gdef\gplfronttext{}%
  \makeatother
  \ifGPblacktext
    \def\colorrgb#1{}%
    \def\colorgray#1{}%
  \else
    \ifGPcolor
      \def\colorrgb#1{\color[rgb]{#1}}%
      \def\colorgray#1{\color[gray]{#1}}%
      \expandafter\def\csname LTw\endcsname{\color{white}}%
      \expandafter\def\csname LTb\endcsname{\color{black}}%
      \expandafter\def\csname LTa\endcsname{\color{black}}%
      \expandafter\def\csname LT0\endcsname{\color[rgb]{1,0,0}}%
      \expandafter\def\csname LT1\endcsname{\color[rgb]{0,1,0}}%
      \expandafter\def\csname LT2\endcsname{\color[rgb]{0,0,1}}%
      \expandafter\def\csname LT3\endcsname{\color[rgb]{1,0,1}}%
      \expandafter\def\csname LT4\endcsname{\color[rgb]{0,1,1}}%
      \expandafter\def\csname LT5\endcsname{\color[rgb]{1,1,0}}%
      \expandafter\def\csname LT6\endcsname{\color[rgb]{0,0,0}}%
      \expandafter\def\csname LT7\endcsname{\color[rgb]{1,0.3,0}}%
      \expandafter\def\csname LT8\endcsname{\color[rgb]{0.5,0.5,0.5}}%
    \else
      \def\colorrgb#1{\color{black}}%
      \def\colorgray#1{\color[gray]{#1}}%
      \expandafter\def\csname LTw\endcsname{\color{white}}%
      \expandafter\def\csname LTb\endcsname{\color{black}}%
      \expandafter\def\csname LTa\endcsname{\color{black}}%
      \expandafter\def\csname LT0\endcsname{\color{black}}%
      \expandafter\def\csname LT1\endcsname{\color{black}}%
      \expandafter\def\csname LT2\endcsname{\color{black}}%
      \expandafter\def\csname LT3\endcsname{\color{black}}%
      \expandafter\def\csname LT4\endcsname{\color{black}}%
      \expandafter\def\csname LT5\endcsname{\color{black}}%
      \expandafter\def\csname LT6\endcsname{\color{black}}%
      \expandafter\def\csname LT7\endcsname{\color{black}}%
      \expandafter\def\csname LT8\endcsname{\color{black}}%
    \fi
  \fi
    \setlength{\unitlength}{0.0500bp}%
    \ifx\gptboxheight\undefined%
      \newlength{\gptboxheight}%
      \newlength{\gptboxwidth}%
      \newsavebox{\gptboxtext}%
    \fi%
    \setlength{\fboxrule}{0.5pt}%
    \setlength{\fboxsep}{1pt}%
\begin{picture}(7200.00,5040.00)%
    \gplgaddtomacro\gplbacktext{%
      \csname LTb\endcsname%
      \put(682,704){\makebox(0,0)[r]{\strut{}$0$}}%
      \put(682,1286){\makebox(0,0)[r]{\strut{}$5$}}%
      \put(682,1867){\makebox(0,0)[r]{\strut{}$10$}}%
      \put(682,2449){\makebox(0,0)[r]{\strut{}$15$}}%
      \put(682,3030){\makebox(0,0)[r]{\strut{}$20$}}%
      \put(682,3612){\makebox(0,0)[r]{\strut{}$25$}}%
      \put(682,4193){\makebox(0,0)[r]{\strut{}$30$}}%
      \put(682,4775){\makebox(0,0)[r]{\strut{}$35$}}%
      \put(814,484){\makebox(0,0){\strut{}$0.4$}}%
      \put(1812,484){\makebox(0,0){\strut{}$0.5$}}%
      \put(2810,484){\makebox(0,0){\strut{}$0.6$}}%
      \put(3808,484){\makebox(0,0){\strut{}$0.7$}}%
      \put(4807,484){\makebox(0,0){\strut{}$0.8$}}%
      \put(5805,484){\makebox(0,0){\strut{}$0.9$}}%
      \put(6803,484){\makebox(0,0){\strut{}$1$}}%
      \put(6000,1030){\makebox(0,0)[l]{\strut{}(b)}}%
    }%
    \gplgaddtomacro\gplfronttext{%
      \csname LTb\endcsname%
      \put(176,2739){\rotatebox{-270}{\makebox(0,0){\strut{}$\mathrm{d}\sigma/\mathrm{d}\cos\theta_{\pi}$  ($10^{-40}~\mathrm{cm}^{2}/\mathrm{nucleon}$)}}}%
      \put(3808,154){\makebox(0,0){\strut{}$\cos\theta_{\pi}$}}%
    }%
    \gplbacktext
    \put(0,0){\includegraphics{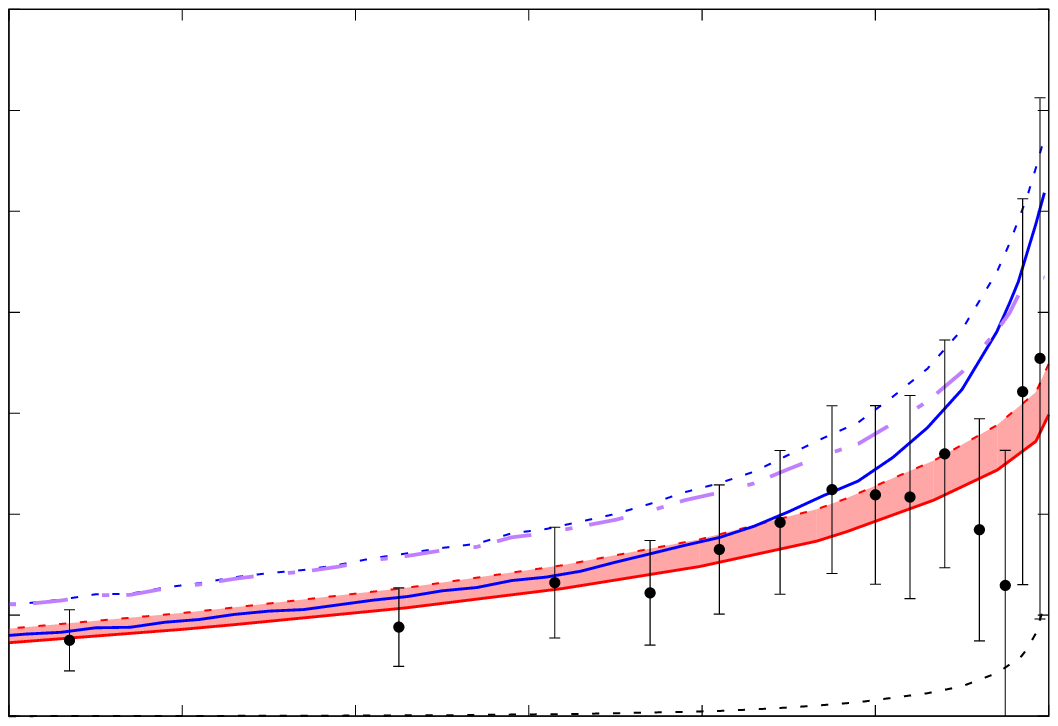}}%
    \gplfronttext
  \end{picture}%
\endgroup

%% file: gfx/CS_NuWro_Pmu_extended.tex
\begingroup
  \makeatletter
  \providecommand\color[2][]{%
    \GenericError{(gnuplot) \space\space\space\@spaces}{%
      Package color not loaded in conjunction with
      terminal option `colourtext'%
    }{See the gnuplot documentation for explanation.%
    }{Either use 'blacktext' in gnuplot or load the package
      color.sty in LaTeX.}%
    \renewcommand\color[2][]{}%
  }%
  \providecommand\includegraphics[2][]{%
    \GenericError{(gnuplot) \space\space\space\@spaces}{%
      Package graphicx or graphics not loaded%
    }{See the gnuplot documentation for explanation.%
    }{The gnuplot epslatex terminal needs graphicx.sty or graphics.sty.}%
    \renewcommand\includegraphics[2][]{}%
  }%
  \providecommand\rotatebox[2]{#2}%
  \@ifundefined{ifGPcolor}{%
    \newif\ifGPcolor
    \GPcolorfalse
  }{}%
  \@ifundefined{ifGPblacktext}{%
    \newif\ifGPblacktext
    \GPblacktexttrue
  }{}%
  \let\gplgaddtomacro\g@addto@macro
  \gdef\gplbacktext{}%
  \gdef\gplfronttext{}%
  \makeatother
  \ifGPblacktext
    \def\colorrgb#1{}%
    \def\colorgray#1{}%
  \else
    \ifGPcolor
      \def\colorrgb#1{\color[rgb]{#1}}%
      \def\colorgray#1{\color[gray]{#1}}%
      \expandafter\def\csname LTw\endcsname{\color{white}}%
      \expandafter\def\csname LTb\endcsname{\color{black}}%
      \expandafter\def\csname LTa\endcsname{\color{black}}%
      \expandafter\def\csname LT0\endcsname{\color[rgb]{1,0,0}}%
      \expandafter\def\csname LT1\endcsname{\color[rgb]{0,1,0}}%
      \expandafter\def\csname LT2\endcsname{\color[rgb]{0,0,1}}%
      \expandafter\def\csname LT3\endcsname{\color[rgb]{1,0,1}}%
      \expandafter\def\csname LT4\endcsname{\color[rgb]{0,1,1}}%
      \expandafter\def\csname LT5\endcsname{\color[rgb]{1,1,0}}%
      \expandafter\def\csname LT6\endcsname{\color[rgb]{0,0,0}}%
      \expandafter\def\csname LT7\endcsname{\color[rgb]{1,0.3,0}}%
      \expandafter\def\csname LT8\endcsname{\color[rgb]{0.5,0.5,0.5}}%
    \else
      \def\colorrgb#1{\color{black}}%
      \def\colorgray#1{\color[gray]{#1}}%
      \expandafter\def\csname LTw\endcsname{\color{white}}%
      \expandafter\def\csname LTb\endcsname{\color{black}}%
      \expandafter\def\csname LTa\endcsname{\color{black}}%
      \expandafter\def\csname LT0\endcsname{\color{black}}%
      \expandafter\def\csname LT1\endcsname{\color{black}}%
      \expandafter\def\csname LT2\endcsname{\color{black}}%
      \expandafter\def\csname LT3\endcsname{\color{black}}%
      \expandafter\def\csname LT4\endcsname{\color{black}}%
      \expandafter\def\csname LT5\endcsname{\color{black}}%
      \expandafter\def\csname LT6\endcsname{\color{black}}%
      \expandafter\def\csname LT7\endcsname{\color{black}}%
      \expandafter\def\csname LT8\endcsname{\color{black}}%
    \fi
  \fi
    \setlength{\unitlength}{0.0500bp}%
    \ifx\gptboxheight\undefined%
      \newlength{\gptboxheight}%
      \newlength{\gptboxwidth}%
      \newsavebox{\gptboxtext}%
    \fi%
    \setlength{\fboxrule}{0.5pt}%
    \setlength{\fboxsep}{1pt}%
\begin{picture}(7200.00,5040.00)%
    \gplgaddtomacro\gplbacktext{%
      \csname LTb\endcsname%
      \put(550,704){\makebox(0,0)[r]{\strut{}$0$}}%
      \put(550,1156){\makebox(0,0)[r]{\strut{}$1$}}%
      \put(550,1609){\makebox(0,0)[r]{\strut{}$2$}}%
      \put(550,2061){\makebox(0,0)[r]{\strut{}$3$}}%
      \put(550,2513){\makebox(0,0)[r]{\strut{}$4$}}%
      \put(550,2966){\makebox(0,0)[r]{\strut{}$5$}}%
      \put(550,3418){\makebox(0,0)[r]{\strut{}$6$}}%
      \put(550,3870){\makebox(0,0)[r]{\strut{}$7$}}%
      \put(550,4323){\makebox(0,0)[r]{\strut{}$8$}}%
      \put(550,4775){\makebox(0,0)[r]{\strut{}$9$}}%
      \put(682,484){\makebox(0,0){\strut{}$0$}}%
      \put(1313,484){\makebox(0,0){\strut{}$0.5$}}%
      \put(1944,484){\makebox(0,0){\strut{}$1$}}%
      \put(2575,484){\makebox(0,0){\strut{}$1.5$}}%
      \put(3206,484){\makebox(0,0){\strut{}$2$}}%
      \put(3837,484){\makebox(0,0){\strut{}$2.5$}}%
      \put(4468,484){\makebox(0,0){\strut{}$3$}}%
      \put(5099,484){\makebox(0,0){\strut{}$3.5$}}%
      \put(5730,484){\makebox(0,0){\strut{}$4$}}%
      \put(6361,484){\makebox(0,0){\strut{}$4.5$}}%
      \put(6191,1111){\makebox(0,0)[l]{\strut{}(a)}}%
    }%
    \gplgaddtomacro\gplfronttext{%
      \csname LTb\endcsname%
      \put(176,2739){\rotatebox{-270}{\makebox(0,0){\strut{}$\mathrm{d}\sigma/\mathrm{d}P_{\mu}$  ($10^{-40}~\mathrm{cm}^{2}/\mathrm{GeV}/\mathrm{nucleon}$)}}}%
      \put(3742,154){\makebox(0,0){\strut{}$P_{\mu}$  (GeV)}}%
      \csname LTb\endcsname%
      \put(5816,4602){\makebox(0,0)[r]{\strut{}Hybrid}}%
      \csname LTb\endcsname%
      \put(5816,4382){\makebox(0,0)[r]{\strut{}Hybrid w/ OSMM}}%
      \csname LTb\endcsname%
      \put(5816,4162){\makebox(0,0)[r]{\strut{}NuWro}}%
      \csname LTb\endcsname%
      \put(5816,3942){\makebox(0,0)[r]{\strut{}NuWro w/o FSI}}%
      \csname LTb\endcsname%
      \put(5816,3722){\makebox(0,0)[r]{\strut{}NuWro w/o FSI $1\pi1N$}}%
      \csname LTb\endcsname%
      \put(5816,3502){\makebox(0,0)[r]{\strut{}NuWro COH}}%
    }%
    \gplbacktext
    \put(0,0){\includegraphics{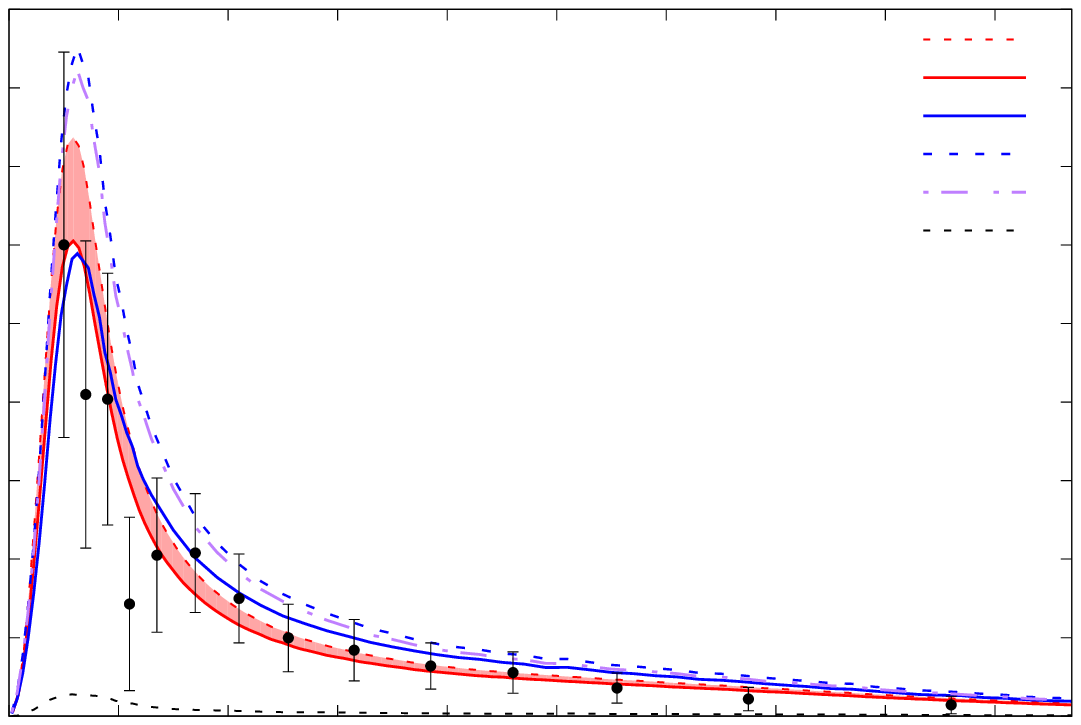}}%
    \gplfronttext
  \end{picture}%
\endgroup

%% file: gfx/CS_NuWro_THl.tex
\begingroup
  \makeatletter
  \providecommand\color[2][]{%
    \GenericError{(gnuplot) \space\space\space\@spaces}{%
      Package color not loaded in conjunction with
      terminal option `colourtext'%
    }{See the gnuplot documentation for explanation.%
    }{Either use 'blacktext' in gnuplot or load the package
      color.sty in LaTeX.}%
    \renewcommand\color[2][]{}%
  }%
  \providecommand\includegraphics[2][]{%
    \GenericError{(gnuplot) \space\space\space\@spaces}{%
      Package graphicx or graphics not loaded%
    }{See the gnuplot documentation for explanation.%
    }{The gnuplot epslatex terminal needs graphicx.sty or graphics.sty.}%
    \renewcommand\includegraphics[2][]{}%
  }%
  \providecommand\rotatebox[2]{#2}%
  \@ifundefined{ifGPcolor}{%
    \newif\ifGPcolor
    \GPcolorfalse
  }{}%
  \@ifundefined{ifGPblacktext}{%
    \newif\ifGPblacktext
    \GPblacktexttrue
  }{}%
  \let\gplgaddtomacro\g@addto@macro
  \gdef\gplbacktext{}%
  \gdef\gplfronttext{}%
  \makeatother
  \ifGPblacktext
    \def\colorrgb#1{}%
    \def\colorgray#1{}%
  \else
    \ifGPcolor
      \def\colorrgb#1{\color[rgb]{#1}}%
      \def\colorgray#1{\color[gray]{#1}}%
      \expandafter\def\csname LTw\endcsname{\color{white}}%
      \expandafter\def\csname LTb\endcsname{\color{black}}%
      \expandafter\def\csname LTa\endcsname{\color{black}}%
      \expandafter\def\csname LT0\endcsname{\color[rgb]{1,0,0}}%
      \expandafter\def\csname LT1\endcsname{\color[rgb]{0,1,0}}%
      \expandafter\def\csname LT2\endcsname{\color[rgb]{0,0,1}}%
      \expandafter\def\csname LT3\endcsname{\color[rgb]{1,0,1}}%
      \expandafter\def\csname LT4\endcsname{\color[rgb]{0,1,1}}%
      \expandafter\def\csname LT5\endcsname{\color[rgb]{1,1,0}}%
      \expandafter\def\csname LT6\endcsname{\color[rgb]{0,0,0}}%
      \expandafter\def\csname LT7\endcsname{\color[rgb]{1,0.3,0}}%
      \expandafter\def\csname LT8\endcsname{\color[rgb]{0.5,0.5,0.5}}%
    \else
      \def\colorrgb#1{\color{black}}%
      \def\colorgray#1{\color[gray]{#1}}%
      \expandafter\def\csname LTw\endcsname{\color{white}}%
      \expandafter\def\csname LTb\endcsname{\color{black}}%
      \expandafter\def\csname LTa\endcsname{\color{black}}%
      \expandafter\def\csname LT0\endcsname{\color{black}}%
      \expandafter\def\csname LT1\endcsname{\color{black}}%
      \expandafter\def\csname LT2\endcsname{\color{black}}%
      \expandafter\def\csname LT3\endcsname{\color{black}}%
      \expandafter\def\csname LT4\endcsname{\color{black}}%
      \expandafter\def\csname LT5\endcsname{\color{black}}%
      \expandafter\def\csname LT6\endcsname{\color{black}}%
      \expandafter\def\csname LT7\endcsname{\color{black}}%
      \expandafter\def\csname LT8\endcsname{\color{black}}%
    \fi
  \fi
    \setlength{\unitlength}{0.0500bp}%
    \ifx\gptboxheight\undefined%
      \newlength{\gptboxheight}%
      \newlength{\gptboxwidth}%
      \newsavebox{\gptboxtext}%
    \fi%
    \setlength{\fboxrule}{0.5pt}%
    \setlength{\fboxsep}{1pt}%
\begin{picture}(7200.00,5040.00)%
    \gplgaddtomacro\gplbacktext{%
      \csname LTb\endcsname%
      \put(682,704){\makebox(0,0)[r]{\strut{}$0$}}%
      \put(682,1156){\makebox(0,0)[r]{\strut{}$10$}}%
      \put(682,1609){\makebox(0,0)[r]{\strut{}$20$}}%
      \put(682,2061){\makebox(0,0)[r]{\strut{}$30$}}%
      \put(682,2513){\makebox(0,0)[r]{\strut{}$40$}}%
      \put(682,2966){\makebox(0,0)[r]{\strut{}$50$}}%
      \put(682,3418){\makebox(0,0)[r]{\strut{}$60$}}%
      \put(682,3870){\makebox(0,0)[r]{\strut{}$70$}}%
      \put(682,4323){\makebox(0,0)[r]{\strut{}$80$}}%
      \put(682,4775){\makebox(0,0)[r]{\strut{}$90$}}%
      \put(1275,484){\makebox(0,0){\strut{}$0.4$}}%
      \put(2196,484){\makebox(0,0){\strut{}$0.5$}}%
      \put(3117,484){\makebox(0,0){\strut{}$0.6$}}%
      \put(4039,484){\makebox(0,0){\strut{}$0.7$}}%
      \put(4960,484){\makebox(0,0){\strut{}$0.8$}}%
      \put(5882,484){\makebox(0,0){\strut{}$0.9$}}%
      \put(6803,484){\makebox(0,0){\strut{}$1$}}%
      \put(6324,1030){\makebox(0,0)[l]{\strut{}(b)}}%
    }%
    \gplgaddtomacro\gplfronttext{%
      \csname LTb\endcsname%
      \put(176,2739){\rotatebox{-270}{\makebox(0,0){\strut{}$\mathrm{d}\sigma/\mathrm{d}\cos\theta_{\mu}$  ($10^{-40}~\mathrm{cm}^{2}/\mathrm{nucleon}$)}}}%
      \put(3808,154){\makebox(0,0){\strut{}$\cos\theta_{\mu}$}}%
    }%
    \gplbacktext
    \put(0,0){\includegraphics{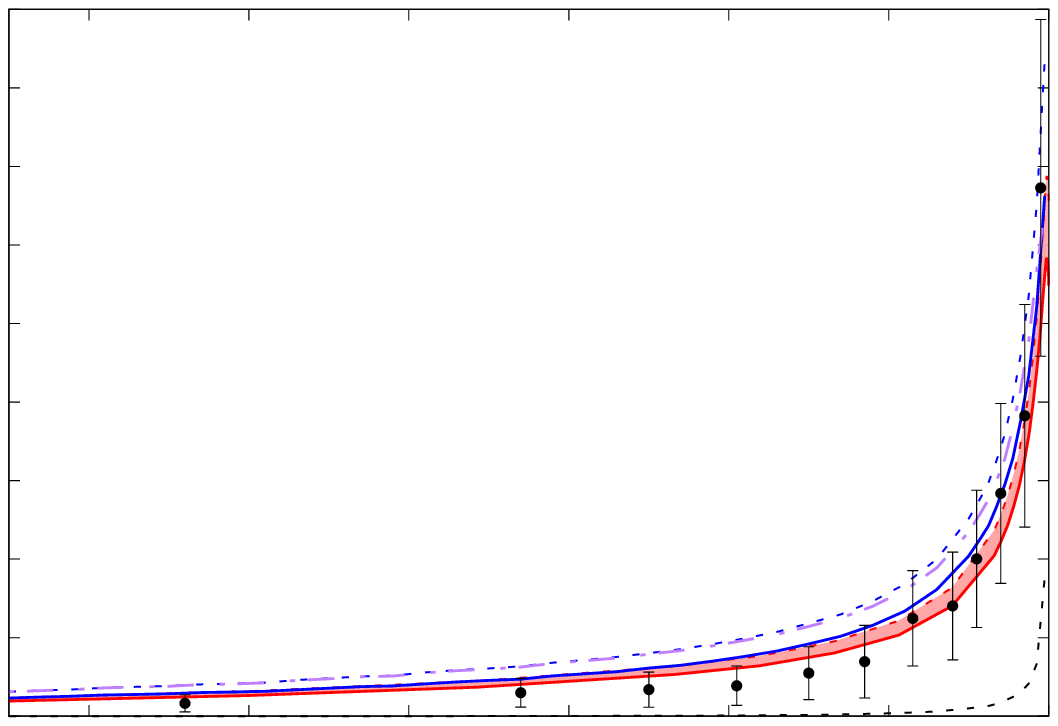}}%
    \gplfronttext
  \end{picture}%
\endgroup

%% file: T2K_H20_Pion.bbl
\begin{thebibliography}{50}%
\makeatletter
\providecommand \@ifxundefined [1]{%
 \@ifx{#1\undefined}
}%
\providecommand \@ifnum [1]{%
 \ifnum #1\expandafter \@firstoftwo
 \else \expandafter \@secondoftwo
 \fi
}%
\providecommand \@ifx [1]{%
 \ifx #1\expandafter \@firstoftwo
 \else \expandafter \@secondoftwo
 \fi
}%
\providecommand \natexlab [1]{#1}%
\providecommand \enquote  [1]{``#1''}%
\providecommand \bibnamefont  [1]{#1}%
\providecommand \bibfnamefont [1]{#1}%
\providecommand \citenamefont [1]{#1}%
\providecommand \href@noop [0]{\@secondoftwo}%
\providecommand \href [0]{\begingroup \@sanitize@url \@href}%
\providecommand \@href[1]{\@@startlink{#1}\@@href}%
\providecommand \@@href[1]{\endgroup#1\@@endlink}%
\providecommand \@sanitize@url [0]{\catcode `\\12\catcode `\$12\catcode
  `\&12\catcode `\#12\catcode `\^12\catcode `\_12\catcode `\%12\relax}%
\providecommand \@@startlink[1]{}%
\providecommand \@@endlink[0]{}%
\providecommand \url  [0]{\begingroup\@sanitize@url \@url }%
\providecommand \@url [1]{\endgroup\@href {#1}{\urlprefix }}%
\providecommand \urlprefix  [0]{URL }%
\providecommand \Eprint [0]{\href }%
\providecommand \doibase [0]{http://dx.doi.org/}%
\providecommand \selectlanguage [0]{\@gobble}%
\providecommand \bibinfo  [0]{\@secondoftwo}%
\providecommand \bibfield  [0]{\@secondoftwo}%
\providecommand \translation [1]{[#1]}%
\providecommand \BibitemOpen [0]{}%
\providecommand \bibitemStop [0]{}%
\providecommand \bibitemNoStop [0]{.\EOS\space}%
\providecommand \EOS [0]{\spacefactor3000\relax}%
\providecommand \BibitemShut  [1]{\csname bibitem#1\endcsname}%
\let\auto@bib@innerbib\@empty
\bibitem [{\citenamefont {Aguilar-Arevalo}\ \emph
  {et~al.}(2011{\natexlab{a}})\citenamefont {Aguilar-Arevalo}, \citenamefont
  {Anderson}, \citenamefont {Bazarko}, \citenamefont {Brice}, \citenamefont
  {Brown} \emph {et~al.}}]{MB:pion}%
  \BibitemOpen
  \bibfield  {author} {\bibinfo {author} {\bibfnamefont {A.~A.}\ \bibnamefont
  {Aguilar-Arevalo}}, \bibinfo {author} {\bibfnamefont {C.~E.}\ \bibnamefont
  {Anderson}}, \bibinfo {author} {\bibfnamefont {A.~O.}\ \bibnamefont
  {Bazarko}}, \bibinfo {author} {\bibfnamefont {S.~J.}\ \bibnamefont {Brice}},
  \bibinfo {author} {\bibfnamefont {B.~C.}\ \bibnamefont {Brown}},  \emph
  {et~al.} (\bibinfo {collaboration} {MiniBooNE Collaboration}),\ }\href
  {\doibase 10.1103/PhysRevD.83.052007} {\bibfield  {journal} {\bibinfo
  {journal} {Phys. Rev. D}\ }\textbf {\bibinfo {volume} {83}},\ \bibinfo
  {pages} {052007} (\bibinfo {year} {2011}{\natexlab{a}})}\BibitemShut
  {NoStop}%
\bibitem [{\citenamefont {Aguilar-Arevalo}\ \emph
  {et~al.}(2011{\natexlab{b}})\citenamefont {Aguilar-Arevalo}, \citenamefont
  {Anderson}, \citenamefont {Bazarko}, \citenamefont {Brice}, \citenamefont
  {Brown}, \citenamefont {Bugel}, \citenamefont {Cao} \emph
  {et~al.}}]{MB:CCneutralpion}%
  \BibitemOpen
  \bibfield  {author} {\bibinfo {author} {\bibfnamefont {A.~A.}\ \bibnamefont
  {Aguilar-Arevalo}}, \bibinfo {author} {\bibfnamefont {C.~E.}\ \bibnamefont
  {Anderson}}, \bibinfo {author} {\bibfnamefont {A.~O.}\ \bibnamefont
  {Bazarko}}, \bibinfo {author} {\bibfnamefont {S.~J.}\ \bibnamefont {Brice}},
  \bibinfo {author} {\bibfnamefont {B.~C.}\ \bibnamefont {Brown}}, \bibinfo
  {author} {\bibfnamefont {L.}~\bibnamefont {Bugel}}, \bibinfo {author}
  {\bibfnamefont {J.}~\bibnamefont {Cao}},  \emph {et~al.} (\bibinfo
  {collaboration} {MiniBooNE Collaboration}),\ }\href {\doibase
  10.1103/PhysRevD.83.052009} {\bibfield  {journal} {\bibinfo  {journal} {Phys.
  Rev. D}\ }\textbf {\bibinfo {volume} {83}},\ \bibinfo {pages} {052009}
  (\bibinfo {year} {2011}{\natexlab{b}})}\BibitemShut {NoStop}%
\bibitem [{\citenamefont {Abe}\ \emph {et~al.}(2013)\citenamefont {Abe},
  \citenamefont {Abgrall}, \citenamefont {Aihara}, \citenamefont {Akiri},
  \citenamefont {Albert}, \citenamefont {Andreopoulos}, \citenamefont {Aoki},
  \citenamefont {Ariga}, \citenamefont {Ariga}, \citenamefont {Assylbekov},
  \citenamefont {Autiero} \emph {et~al.}}]{T2K:Inclnumu}%
  \BibitemOpen
  \bibfield  {author} {\bibinfo {author} {\bibfnamefont {K.}~\bibnamefont
  {Abe}}, \bibinfo {author} {\bibfnamefont {N.}~\bibnamefont {Abgrall}},
  \bibinfo {author} {\bibfnamefont {H.}~\bibnamefont {Aihara}}, \bibinfo
  {author} {\bibfnamefont {T.}~\bibnamefont {Akiri}}, \bibinfo {author}
  {\bibfnamefont {J.~B.}\ \bibnamefont {Albert}}, \bibinfo {author}
  {\bibfnamefont {C.}~\bibnamefont {Andreopoulos}}, \bibinfo {author}
  {\bibfnamefont {S.}~\bibnamefont {Aoki}}, \bibinfo {author} {\bibfnamefont
  {A.}~\bibnamefont {Ariga}}, \bibinfo {author} {\bibfnamefont
  {T.}~\bibnamefont {Ariga}}, \bibinfo {author} {\bibfnamefont
  {S.}~\bibnamefont {Assylbekov}}, \bibinfo {author} {\bibfnamefont
  {D.}~\bibnamefont {Autiero}},  \emph {et~al.} (\bibinfo {collaboration} {T2K
  Collaboration}),\ }\href {\doibase 10.1103/PhysRevD.87.092003} {\bibfield
  {journal} {\bibinfo  {journal} {Phys. Rev. D}\ }\textbf {\bibinfo {volume}
  {87}},\ \bibinfo {pages} {092003} (\bibinfo {year} {2013})}\BibitemShut
  {NoStop}%
\bibitem [{\citenamefont {Abe}\ \emph {et~al.}(2017)\citenamefont {Abe},
  \citenamefont {Andreopoulos}, \citenamefont {Antonova}, \citenamefont {Aoki},
  \citenamefont {Ariga}, \citenamefont {Assylbekov}, \citenamefont {Autiero},
  \citenamefont {Ban}, \citenamefont {Barbi}, \citenamefont {Barker} \emph
  {et~al.}}]{T2KCC1PIH2O}%
  \BibitemOpen
  \bibfield  {author} {\bibinfo {author} {\bibfnamefont {K.}~\bibnamefont
  {Abe}}, \bibinfo {author} {\bibfnamefont {C.}~\bibnamefont {Andreopoulos}},
  \bibinfo {author} {\bibfnamefont {M.}~\bibnamefont {Antonova}}, \bibinfo
  {author} {\bibfnamefont {S.}~\bibnamefont {Aoki}}, \bibinfo {author}
  {\bibfnamefont {A.}~\bibnamefont {Ariga}}, \bibinfo {author} {\bibfnamefont
  {S.}~\bibnamefont {Assylbekov}}, \bibinfo {author} {\bibfnamefont
  {D.}~\bibnamefont {Autiero}}, \bibinfo {author} {\bibfnamefont
  {S.}~\bibnamefont {Ban}}, \bibinfo {author} {\bibfnamefont {M.}~\bibnamefont
  {Barbi}}, \bibinfo {author} {\bibfnamefont {G.~J.}\ \bibnamefont {Barker}},
  \emph {et~al.} (\bibinfo {collaboration} {T2K Collaboration}),\ }\href
  {\doibase 10.1103/PhysRevD.95.012010} {\bibfield  {journal} {\bibinfo
  {journal} {Phys. Rev. D}\ }\textbf {\bibinfo {volume} {95}},\ \bibinfo
  {pages} {012010} (\bibinfo {year} {2017})}\BibitemShut {NoStop}%
\bibitem [{\citenamefont {Eberly}\ \emph {et~al.}(2015)\citenamefont {Eberly},
  \citenamefont {Aliaga}, \citenamefont {Altinok}, \citenamefont
  {Barrios~Sazo}, \citenamefont {Bellantoni}, \citenamefont {Betancourt} \emph
  {et~al.}}]{MINERvA:CC1PI}%
  \BibitemOpen
  \bibfield  {author} {\bibinfo {author} {\bibfnamefont {B.}~\bibnamefont
  {Eberly}}, \bibinfo {author} {\bibfnamefont {L.}~\bibnamefont {Aliaga}},
  \bibinfo {author} {\bibfnamefont {O.}~\bibnamefont {Altinok}}, \bibinfo
  {author} {\bibfnamefont {M.~G.}\ \bibnamefont {Barrios~Sazo}}, \bibinfo
  {author} {\bibfnamefont {L.}~\bibnamefont {Bellantoni}}, \bibinfo {author}
  {\bibfnamefont {M.}~\bibnamefont {Betancourt}},  \emph {et~al.} (\bibinfo
  {collaboration} {MINERvA Collaboration}),\ }\href {\doibase
  10.1103/PhysRevD.92.092008} {\bibfield  {journal} {\bibinfo  {journal} {Phys.
  Rev. D}\ }\textbf {\bibinfo {volume} {92}},\ \bibinfo {pages} {092008}
  (\bibinfo {year} {2015})}\BibitemShut {NoStop}%
\bibitem [{\citenamefont {Altinok}\ \emph {et~al.}(2017)\citenamefont
  {Altinok}, \citenamefont {Le}, \citenamefont {Aliaga}, \citenamefont
  {Bellantoni}, \citenamefont {Bercellie}, \citenamefont {Betancourt} \emph
  {et~al.}}]{MINERvA:CCPI0}%
  \BibitemOpen
  \bibfield  {author} {\bibinfo {author} {\bibfnamefont {O.}~\bibnamefont
  {Altinok}}, \bibinfo {author} {\bibfnamefont {T.}~\bibnamefont {Le}},
  \bibinfo {author} {\bibfnamefont {L.}~\bibnamefont {Aliaga}}, \bibinfo
  {author} {\bibfnamefont {L.}~\bibnamefont {Bellantoni}}, \bibinfo {author}
  {\bibfnamefont {A.}~\bibnamefont {Bercellie}}, \bibinfo {author}
  {\bibfnamefont {M.}~\bibnamefont {Betancourt}},  \emph {et~al.},\ }\href
  {\doibase 10.1103/PhysRevD.96.072003} {\bibfield  {journal} {\bibinfo
  {journal} {Phys. Rev. D}\ }\textbf {\bibinfo {volume} {96}},\ \bibinfo
  {pages} {072003} (\bibinfo {year} {2017})}\BibitemShut {NoStop}%
\bibitem [{NOV()}]{NOVA}%
  \BibitemOpen
  \href {https://www-nova.fnal.gov/} {}\bibinfo {howpublished}
  {\url{https://www-nova.fnal.gov/}}\BibitemShut {NoStop}%
\bibitem [{DUN()}]{DUNE}%
  \BibitemOpen
  \href {http://www.dunescience.org} {}\bibinfo {howpublished}
  {\url{http://www.dunescience.org}}\BibitemShut {NoStop}%
\bibitem [{\citenamefont {Alvarez-Ruso}\ \emph {et~al.}(2018)\citenamefont
  {Alvarez-Ruso}, \citenamefont {Sajjad Athar}, \citenamefont {Barbaro},
  \citenamefont {Cherdack}, \citenamefont {Christy}, \citenamefont {Coloma},
  \citenamefont {Donnelly}, \citenamefont {Dytman}, \citenamefont
  {de Gouv\^{e}a}, \citenamefont {Hill}, \citenamefont {Huber}, \citenamefont
  {Jachowicz}, \citenamefont {Katori}, \citenamefont {Kronfeld}, \citenamefont
  {Mahn}, \citenamefont {Martini}, \citenamefont {Morf\'{i}n}, \citenamefont
  {Nieves}, \citenamefont {Perdue}, \citenamefont {Petti}, \citenamefont
  {Richards}, \citenamefont {S\'{a}nchez}, \citenamefont {Sato}, \citenamefont
  {Sobczyk},\ and\ \citenamefont {Zeller}}]{NUSTECWP}%
  \BibitemOpen
  \bibfield  {author} {\bibinfo {author} {\bibfnamefont {L.}~\bibnamefont
  {Alvarez-Ruso}}, \bibinfo {author} {\bibfnamefont {M.}~\bibnamefont
  {Sajjad Athar}}, \bibinfo {author} {\bibfnamefont {M.}~\bibnamefont
  {Barbaro}}, \bibinfo {author} {\bibfnamefont {D.}~\bibnamefont {Cherdack}},
  \bibinfo {author} {\bibfnamefont {M.}~\bibnamefont {Christy}}, \bibinfo
  {author} {\bibfnamefont {P.}~\bibnamefont {Coloma}}, \bibinfo {author}
  {\bibfnamefont {T.}~\bibnamefont {Donnelly}}, \bibinfo {author}
  {\bibfnamefont {S.}~\bibnamefont {Dytman}}, \bibinfo {author} {\bibfnamefont
  {A.}~\bibnamefont {de Gouv\^{e}a}}, \bibinfo {author} {\bibfnamefont
  {R.}~\bibnamefont {Hill}}, \bibinfo {author} {\bibfnamefont {P.}~\bibnamefont
  {Huber}}, \bibinfo {author} {\bibfnamefont {N.}~\bibnamefont {Jachowicz}},
  \bibinfo {author} {\bibfnamefont {T.}~\bibnamefont {Katori}}, \bibinfo
  {author} {\bibfnamefont {A.}~\bibnamefont {Kronfeld}}, \bibinfo {author}
  {\bibfnamefont {K.}~\bibnamefont {Mahn}}, \bibinfo {author} {\bibfnamefont
  {M.}~\bibnamefont {Martini}}, \bibinfo {author} {\bibfnamefont
  {J.}~\bibnamefont {Morf\'{i}n}}, \bibinfo {author} {\bibfnamefont
  {J.}~\bibnamefont {Nieves}}, \bibinfo {author} {\bibfnamefont
  {G.}~\bibnamefont {Perdue}}, \bibinfo {author} {\bibfnamefont
  {R.}~\bibnamefont {Petti}}, \bibinfo {author} {\bibfnamefont
  {D.}~\bibnamefont {Richards}}, \bibinfo {author} {\bibfnamefont
  {F.}~\bibnamefont {S\'{a}nchez}}, \bibinfo {author} {\bibfnamefont
  {T.}~\bibnamefont {Sato}}, \bibinfo {author} {\bibfnamefont {J.}~\bibnamefont
  {Sobczyk}}, \ and\ \bibinfo {author} {\bibfnamefont {G.}~\bibnamefont
  {Zeller}},\ }\href
  {http://www.sciencedirect.com/science/article/pii/S0146641018300061}
  {\bibfield  {journal} {\bibinfo  {journal} {Progress in Particle and Nuclear
  Physics}\ } (\bibinfo {year} {2018})}\BibitemShut {NoStop}%
\bibitem [{\citenamefont {Martini}\ \emph {et~al.}(2009)\citenamefont
  {Martini}, \citenamefont {Ericson}, \citenamefont {Chanfray},\ and\
  \citenamefont {Marteau}}]{MartiniModel:2009}%
  \BibitemOpen
  \bibfield  {author} {\bibinfo {author} {\bibfnamefont {M.}~\bibnamefont
  {Martini}}, \bibinfo {author} {\bibfnamefont {M.}~\bibnamefont {Ericson}},
  \bibinfo {author} {\bibfnamefont {G.}~\bibnamefont {Chanfray}}, \ and\
  \bibinfo {author} {\bibfnamefont {J.}~\bibnamefont {Marteau}},\ }\href
  {\doibase 10.1103/PhysRevC.80.065501} {\bibfield  {journal} {\bibinfo
  {journal} {Phys. Rev. C}\ }\textbf {\bibinfo {volume} {80}},\ \bibinfo
  {pages} {065501} (\bibinfo {year} {2009})}\BibitemShut {NoStop}%
\bibitem [{\citenamefont {Rein}\ and\ \citenamefont
  {Sehgal}(1981)}]{ReinSeghal}%
  \BibitemOpen
  \bibfield  {author} {\bibinfo {author} {\bibfnamefont {D.}~\bibnamefont
  {Rein}}\ and\ \bibinfo {author} {\bibfnamefont {L.~M.}\ \bibnamefont
  {Sehgal}},\ }\href@noop {} {\bibfield  {journal} {\bibinfo  {journal} {Annals
  of Physics}\ }\textbf {\bibinfo {volume} {133}},\ \bibinfo {pages} {97}
  (\bibinfo {year} {1981})}\BibitemShut {NoStop}%
\bibitem [{\citenamefont {Hern\'andez}\ \emph {et~al.}(2007)\citenamefont
  {Hern\'andez}, \citenamefont {Nieves},\ and\ \citenamefont
  {Valverde}}]{Hernandez:Pion}%
  \BibitemOpen
  \bibfield  {author} {\bibinfo {author} {\bibfnamefont {E.}~\bibnamefont
  {Hern\'andez}}, \bibinfo {author} {\bibfnamefont {J.}~\bibnamefont {Nieves}},
  \ and\ \bibinfo {author} {\bibfnamefont {M.}~\bibnamefont {Valverde}},\
  }\href {\doibase 10.1103/PhysRevD.76.033005} {\bibfield  {journal} {\bibinfo
  {journal} {Phys. Rev. D}\ }\textbf {\bibinfo {volume} {76}},\ \bibinfo
  {pages} {033005} (\bibinfo {year} {2007})}\BibitemShut {NoStop}%
\bibitem [{\citenamefont {Yu}\ \emph {et~al.}(2015)\citenamefont {Yu},
  \citenamefont {Paschos},\ and\ \citenamefont {Schienbein}}]{Paschos:2015}%
  \BibitemOpen
  \bibfield  {author} {\bibinfo {author} {\bibfnamefont {J.~Y.}\ \bibnamefont
  {Yu}}, \bibinfo {author} {\bibfnamefont {E.~A.}\ \bibnamefont {Paschos}}, \
  and\ \bibinfo {author} {\bibfnamefont {I.}~\bibnamefont {Schienbein}},\
  }\href {\doibase 10.1103/PhysRevD.91.054038} {\bibfield  {journal} {\bibinfo
  {journal} {Phys. Rev. D}\ }\textbf {\bibinfo {volume} {91}},\ \bibinfo
  {pages} {054038} (\bibinfo {year} {2015})}\BibitemShut {NoStop}%
\bibitem [{\citenamefont {Ivanov}\ \emph {et~al.}(2016)\citenamefont {Ivanov},
  \citenamefont {Megias}, \citenamefont {Gonz\'{a}lez-Jim\'{e}nez},
  \citenamefont {Moreno}, \citenamefont {Barbaro}, \citenamefont {Caballero},\
  and\ \citenamefont {Donnelly}}]{SUSA:2016}%
  \BibitemOpen
  \bibfield  {author} {\bibinfo {author} {\bibfnamefont {M.~V.}\ \bibnamefont
  {Ivanov}}, \bibinfo {author} {\bibfnamefont {G.~D.}\ \bibnamefont {Megias}},
  \bibinfo {author} {\bibfnamefont {R.}~\bibnamefont
  {Gonz\'{a}lez-Jim\'{e}nez}}, \bibinfo {author} {\bibfnamefont
  {O.}~\bibnamefont {Moreno}}, \bibinfo {author} {\bibfnamefont {M.~B.}\
  \bibnamefont {Barbaro}}, \bibinfo {author} {\bibfnamefont {J.~A.}\
  \bibnamefont {Caballero}}, \ and\ \bibinfo {author} {\bibfnamefont {T.~W.}\
  \bibnamefont {Donnelly}},\ }\href
  {http://stacks.iop.org/0954-3899/43/i=4/a=045101} {\bibfield  {journal}
  {\bibinfo  {journal} {Journal of Physics G}\ }\textbf {\bibinfo {volume}
  {43}},\ \bibinfo {pages} {045101} (\bibinfo {year} {2016})}\BibitemShut
  {NoStop}%
\bibitem [{\citenamefont {Buss}\ \emph {et~al.}(2007)\citenamefont {Buss},
  \citenamefont {Leitner}, \citenamefont {Mosel},\ and\ \citenamefont
  {Alvarez-Ruso}}]{BussMosel}%
  \BibitemOpen
  \bibfield  {author} {\bibinfo {author} {\bibfnamefont {O.}~\bibnamefont
  {Buss}}, \bibinfo {author} {\bibfnamefont {T.}~\bibnamefont {Leitner}},
  \bibinfo {author} {\bibfnamefont {U.}~\bibnamefont {Mosel}}, \ and\ \bibinfo
  {author} {\bibfnamefont {L.}~\bibnamefont {Alvarez-Ruso}},\ }\href {\doibase
  10.1103/PhysRevC.76.035502} {\bibfield  {journal} {\bibinfo  {journal} {Phys.
  Rev. C}\ }\textbf {\bibinfo {volume} {76}},\ \bibinfo {pages} {035502}
  (\bibinfo {year} {2007})}\BibitemShut {NoStop}%
\bibitem [{\citenamefont {Praet}\ \emph {et~al.}(2009)\citenamefont {Praet},
  \citenamefont {Lalakulich}, \citenamefont {Jachowicz},\ and\ \citenamefont
  {Ryckebusch}}]{Praet}%
  \BibitemOpen
  \bibfield  {author} {\bibinfo {author} {\bibfnamefont {C.}~\bibnamefont
  {Praet}}, \bibinfo {author} {\bibfnamefont {O.}~\bibnamefont {Lalakulich}},
  \bibinfo {author} {\bibfnamefont {N.}~\bibnamefont {Jachowicz}}, \ and\
  \bibinfo {author} {\bibfnamefont {J.}~\bibnamefont {Ryckebusch}},\ }\href
  {\doibase 10.1103/PhysRevC.79.044603} {\bibfield  {journal} {\bibinfo
  {journal} {Phys. Rev. C}\ }\textbf {\bibinfo {volume} {79}},\ \bibinfo
  {pages} {044603} (\bibinfo {year} {2009})}\BibitemShut {NoStop}%
\bibitem [{\citenamefont {Ahmad}\ \emph {et~al.}(2006)\citenamefont {Ahmad},
  \citenamefont {Athar},\ and\ \citenamefont {Singh}}]{Singh:2006}%
  \BibitemOpen
  \bibfield  {author} {\bibinfo {author} {\bibfnamefont {S.}~\bibnamefont
  {Ahmad}}, \bibinfo {author} {\bibfnamefont {M.~S.}\ \bibnamefont {Athar}}, \
  and\ \bibinfo {author} {\bibfnamefont {S.~K.}\ \bibnamefont {Singh}},\ }\href
  {\doibase 10.1103/PhysRevD.74.073008} {\bibfield  {journal} {\bibinfo
  {journal} {Phys. Rev. D}\ }\textbf {\bibinfo {volume} {74}},\ \bibinfo
  {pages} {073008} (\bibinfo {year} {2006})}\BibitemShut {NoStop}%
\bibitem [{\citenamefont {Rafi~Alam}\ \emph {et~al.}(2016)\citenamefont
  {Rafi~Alam}, \citenamefont {Sajjad~Athar}, \citenamefont {Chauhan},\ and\
  \citenamefont {Singh}}]{Singh:2016}%
  \BibitemOpen
  \bibfield  {author} {\bibinfo {author} {\bibfnamefont {M.}~\bibnamefont
  {Rafi~Alam}}, \bibinfo {author} {\bibfnamefont {M.}~\bibnamefont
  {Sajjad~Athar}}, \bibinfo {author} {\bibfnamefont {S.}~\bibnamefont
  {Chauhan}}, \ and\ \bibinfo {author} {\bibfnamefont {S.~K.}\ \bibnamefont
  {Singh}},\ }\href {\doibase 10.1142/S0218301316500105} {\bibfield  {journal}
  {\bibinfo  {journal} {International Journal of Modern Physics E}\ }\textbf
  {\bibinfo {volume} {25}},\ \bibinfo {pages} {1650010} (\bibinfo {year}
  {2016})}\BibitemShut {NoStop}%
\bibitem [{\citenamefont {Zhang}\ and\ \citenamefont
  {Serot}(2012)}]{ZhangSerot}%
  \BibitemOpen
  \bibfield  {author} {\bibinfo {author} {\bibfnamefont {X.}~\bibnamefont
  {Zhang}}\ and\ \bibinfo {author} {\bibfnamefont {B.~D.}\ \bibnamefont
  {Serot}},\ }\href {\doibase 10.1103/PhysRevC.86.035504} {\bibfield  {journal}
  {\bibinfo  {journal} {Phys. Rev. C}\ }\textbf {\bibinfo {volume} {86}},\
  \bibinfo {pages} {035504} (\bibinfo {year} {2012})}\BibitemShut {NoStop}%
\bibitem [{\citenamefont {Lalakulich}\ and\ \citenamefont
  {Mosel}(2013)}]{Mosel:MB}%
  \BibitemOpen
  \bibfield  {author} {\bibinfo {author} {\bibfnamefont {O.}~\bibnamefont
  {Lalakulich}}\ and\ \bibinfo {author} {\bibfnamefont {U.}~\bibnamefont
  {Mosel}},\ }\href {\doibase 10.1103/PhysRevC.87.014602} {\bibfield  {journal}
  {\bibinfo  {journal} {Phys. Rev. C}\ }\textbf {\bibinfo {volume} {87}},\
  \bibinfo {pages} {014602} (\bibinfo {year} {2013})}\BibitemShut {NoStop}%
\bibitem [{\citenamefont {Nakamura}\ \emph {et~al.}(2015)\citenamefont
  {Nakamura}, \citenamefont {Kamano},\ and\ \citenamefont {Sato}}]{Sato}%
  \BibitemOpen
  \bibfield  {author} {\bibinfo {author} {\bibfnamefont {S.~X.}\ \bibnamefont
  {Nakamura}}, \bibinfo {author} {\bibfnamefont {H.}~\bibnamefont {Kamano}}, \
  and\ \bibinfo {author} {\bibfnamefont {T.}~\bibnamefont {Sato}},\ }\href
  {\doibase 10.1103/PhysRevD.92.074024} {\bibfield  {journal} {\bibinfo
  {journal} {Phys. Rev. D}\ }\textbf {\bibinfo {volume} {92}},\ \bibinfo
  {pages} {074024} (\bibinfo {year} {2015})}\BibitemShut {NoStop}%
\bibitem [{\citenamefont {Hern\'andez}\ \emph {et~al.}(2013)\citenamefont
  {Hern\'andez}, \citenamefont {Nieves},\ and\ \citenamefont
  {Vacas}}]{Hernandez:PionNucleus}%
  \BibitemOpen
  \bibfield  {author} {\bibinfo {author} {\bibfnamefont {E.}~\bibnamefont
  {Hern\'andez}}, \bibinfo {author} {\bibfnamefont {J.}~\bibnamefont {Nieves}},
  \ and\ \bibinfo {author} {\bibfnamefont {M.~J.~V.}\ \bibnamefont {Vacas}},\
  }\href {\doibase 10.1103/PhysRevD.87.113009} {\bibfield  {journal} {\bibinfo
  {journal} {Phys. Rev. D}\ }\textbf {\bibinfo {volume} {87}},\ \bibinfo
  {pages} {113009} (\bibinfo {year} {2013})}\BibitemShut {NoStop}%
\bibitem [{\citenamefont {Gonz\'{a}lez-Jim\'{e}nez}\ \emph
  {et~al.}(2017)\citenamefont {Gonz\'{a}lez-Jim\'{e}nez}, \citenamefont
  {Jachowicz}, \citenamefont {Niewczas}, \citenamefont {Nys}, \citenamefont
  {Pandey}, \citenamefont {Van~Cuyck},\ and\ \citenamefont
  {Van~Dessel}}]{Gonzalez:SPPnucleon}%
  \BibitemOpen
  \bibfield  {author} {\bibinfo {author} {\bibfnamefont {R.}~\bibnamefont
  {Gonz\'{a}lez-Jim\'{e}nez}}, \bibinfo {author} {\bibfnamefont
  {N.}~\bibnamefont {Jachowicz}}, \bibinfo {author} {\bibfnamefont
  {K.}~\bibnamefont {Niewczas}}, \bibinfo {author} {\bibfnamefont
  {J.}~\bibnamefont {Nys}}, \bibinfo {author} {\bibfnamefont {V.}~\bibnamefont
  {Pandey}}, \bibinfo {author} {\bibfnamefont {T.}~\bibnamefont {Van~Cuyck}}, \
  and\ \bibinfo {author} {\bibfnamefont {N.}~\bibnamefont {Van~Dessel}},\
  }\href {\doibase 10.1103/PhysRevD.95.113007} {\bibfield  {journal} {\bibinfo
  {journal} {Phys. Rev. D}\ }\textbf {\bibinfo {volume} {95}},\ \bibinfo
  {pages} {113007} (\bibinfo {year} {2017})}\BibitemShut {NoStop}%
\bibitem [{\citenamefont {Scherer}\ and\ \citenamefont
  {Schindler}(2012)}]{ChPT}%
  \BibitemOpen
  \bibfield  {author} {\bibinfo {author} {\bibfnamefont {S.}~\bibnamefont
  {Scherer}}\ and\ \bibinfo {author} {\bibfnamefont {M.~R.}\ \bibnamefont
  {Schindler}},\ }\href@noop {} {\emph {\bibinfo {title} {A primer for Chiral
  perturbation theory}}}\ (\bibinfo  {publisher} {Springer},\ \bibinfo {year}
  {2012})\BibitemShut {NoStop}%
\bibitem [{\citenamefont {Lalakulich}\ \emph {et~al.}(2006)\citenamefont
  {Lalakulich}, \citenamefont {Paschos},\ and\ \citenamefont
  {Piranishvili}}]{Lalakulich:Res}%
  \BibitemOpen
  \bibfield  {author} {\bibinfo {author} {\bibfnamefont {O.}~\bibnamefont
  {Lalakulich}}, \bibinfo {author} {\bibfnamefont {E.~A.}\ \bibnamefont
  {Paschos}}, \ and\ \bibinfo {author} {\bibfnamefont {G.}~\bibnamefont
  {Piranishvili}},\ }\href {\doibase 10.1103/PhysRevD.74.014009} {\bibfield
  {journal} {\bibinfo  {journal} {Phys. Rev. D}\ }\textbf {\bibinfo {volume}
  {74}},\ \bibinfo {pages} {014009} (\bibinfo {year} {2006})}\BibitemShut
  {NoStop}%
\bibitem [{\citenamefont {Davidson}\ and\ \citenamefont
  {Workman}(2001)}]{FFres}%
  \BibitemOpen
  \bibfield  {author} {\bibinfo {author} {\bibfnamefont {R.~M.}\ \bibnamefont
  {Davidson}}\ and\ \bibinfo {author} {\bibfnamefont {R.}~\bibnamefont
  {Workman}},\ }\href {\doibase 10.1103/PhysRevC.63.025210} {\bibfield
  {journal} {\bibinfo  {journal} {Phys. Rev. C}\ }\textbf {\bibinfo {volume}
  {63}},\ \bibinfo {pages} {025210} (\bibinfo {year} {2001})}\BibitemShut
  {NoStop}%
\bibitem [{\citenamefont {Vrancx}\ \emph {et~al.}(2011)\citenamefont {Vrancx},
  \citenamefont {De~Cruz}, \citenamefont {Ryckebusch},\ and\ \citenamefont
  {Vancraeyveld}}]{Vrancx}%
  \BibitemOpen
  \bibfield  {author} {\bibinfo {author} {\bibfnamefont {T.}~\bibnamefont
  {Vrancx}}, \bibinfo {author} {\bibfnamefont {L.}~\bibnamefont {De~Cruz}},
  \bibinfo {author} {\bibfnamefont {J.}~\bibnamefont {Ryckebusch}}, \ and\
  \bibinfo {author} {\bibfnamefont {P.}~\bibnamefont {Vancraeyveld}},\ }\href
  {\doibase 10.1103/PhysRevC.84.045201} {\bibfield  {journal} {\bibinfo
  {journal} {Phys. Rev. C}\ }\textbf {\bibinfo {volume} {84}},\ \bibinfo
  {pages} {045201} (\bibinfo {year} {2011})}\BibitemShut {NoStop}%
\bibitem [{\citenamefont {Guidal}\ \emph {et~al.}(1997)\citenamefont {Guidal},
  \citenamefont {Laget},\ and\ \citenamefont {Vanderhaeghen}}]{GVL}%
  \BibitemOpen
  \bibfield  {author} {\bibinfo {author} {\bibfnamefont {M.}~\bibnamefont
  {Guidal}}, \bibinfo {author} {\bibfnamefont {J.-M.}\ \bibnamefont {Laget}}, \
  and\ \bibinfo {author} {\bibfnamefont {M.}~\bibnamefont {Vanderhaeghen}},\
  }\href {\doibase https://doi.org/10.1016/S0375-9474(97)00612-X} {\bibfield
  {journal} {\bibinfo  {journal} {Nuclear Physics A}\ }\textbf {\bibinfo
  {volume} {627}},\ \bibinfo {pages} {645 } (\bibinfo {year}
  {1997})}\BibitemShut {NoStop}%
\bibitem [{\citenamefont {Kaskulov}\ and\ \citenamefont
  {Mosel}(2010)}]{Kaskulov:Mosel}%
  \BibitemOpen
  \bibfield  {author} {\bibinfo {author} {\bibfnamefont {M.~M.}\ \bibnamefont
  {Kaskulov}}\ and\ \bibinfo {author} {\bibfnamefont {U.}~\bibnamefont
  {Mosel}},\ }\href {\doibase 10.1103/PhysRevC.81.045202} {\bibfield  {journal}
  {\bibinfo  {journal} {Phys. Rev. C}\ }\textbf {\bibinfo {volume} {81}},\
  \bibinfo {pages} {045202} (\bibinfo {year} {2010})}\BibitemShut {NoStop}%
\bibitem [{\citenamefont {Gonz\'alez-Jim\'enez}\ \emph
  {et~al.}(2018)\citenamefont {Gonz\'alez-Jim\'enez}, \citenamefont
  {Niewczas},\ and\ \citenamefont {Jachowicz}}]{HybridRPWIA}%
  \BibitemOpen
  \bibfield  {author} {\bibinfo {author} {\bibfnamefont {R.}~\bibnamefont
  {Gonz\'alez-Jim\'enez}}, \bibinfo {author} {\bibfnamefont {K.}~\bibnamefont
  {Niewczas}}, \ and\ \bibinfo {author} {\bibfnamefont {N.}~\bibnamefont
  {Jachowicz}},\ }\href {\doibase 10.1103/PhysRevD.97.013004} {\bibfield
  {journal} {\bibinfo  {journal} {Phys. Rev. D}\ }\textbf {\bibinfo {volume}
  {97}},\ \bibinfo {pages} {013004} (\bibinfo {year} {2018})}\BibitemShut
  {NoStop}%
\bibitem [{\citenamefont {Ring}(1996)}]{RINGRMF}%
  \BibitemOpen
  \bibfield  {author} {\bibinfo {author} {\bibfnamefont {P.}~\bibnamefont
  {Ring}},\ }\href {\doibase https://doi.org/10.1016/0146-6410(96)00054-3}
  {\bibfield  {journal} {\bibinfo  {journal} {Progress in Particle and Nuclear
  Physics}\ }\textbf {\bibinfo {volume} {37}},\ \bibinfo {pages} {193 }
  (\bibinfo {year} {1996})}\BibitemShut {NoStop}%
\bibitem [{\citenamefont {Walecka}(1974)}]{WALECKARMF}%
  \BibitemOpen
  \bibfield  {author} {\bibinfo {author} {\bibfnamefont {J.}~\bibnamefont
  {Walecka}},\ }\href {\doibase https://doi.org/10.1016/0003-4916(74)90208-5}
  {\bibfield  {journal} {\bibinfo  {journal} {Annals of Physics}\ }\textbf
  {\bibinfo {volume} {83}},\ \bibinfo {pages} {491 } (\bibinfo {year}
  {1974})}\BibitemShut {NoStop}%
\bibitem [{\citenamefont {Hayato}(2009)}]{Hayato:NEUT}%
  \BibitemOpen
  \bibfield  {author} {\bibinfo {author} {\bibfnamefont {Y.}~\bibnamefont
  {Hayato}},\ }\href@noop {} {\bibfield  {journal} {\bibinfo  {journal} {Acta
  Phys. Polon.}\ }\textbf {\bibinfo {volume} {B40}},\ \bibinfo {pages} {2477}
  (\bibinfo {year} {2009})}\BibitemShut {NoStop}%
\bibitem [{\citenamefont {Andreopoulos}\ \emph {et~al.}(2010)\citenamefont
  {Andreopoulos} \emph {et~al.}}]{GENIE}%
  \BibitemOpen
  \bibfield  {author} {\bibinfo {author} {\bibfnamefont {C.}~\bibnamefont
  {Andreopoulos}} \emph {et~al.},\ }\href {\doibase 10.1016/j.nima.2009.12.009}
  {\bibfield  {journal} {\bibinfo  {journal} {Nucl. Instrum. Meth.}\ }\textbf
  {\bibinfo {volume} {A614}},\ \bibinfo {pages} {87} (\bibinfo {year}
  {2010})},\ \Eprint {http://arxiv.org/abs/0905.2517} {arXiv:0905.2517
  [hep-ph]} \BibitemShut {NoStop}%
\bibitem [{\citenamefont {Golan}\ \emph {et~al.}(2012)\citenamefont {Golan},
  \citenamefont {Juszczak},\ and\ \citenamefont {Sobczyk}}]{NuWroFSI}%
  \BibitemOpen
  \bibfield  {author} {\bibinfo {author} {\bibfnamefont {T.}~\bibnamefont
  {Golan}}, \bibinfo {author} {\bibfnamefont {C.}~\bibnamefont {Juszczak}}, \
  and\ \bibinfo {author} {\bibfnamefont {J.~T.}\ \bibnamefont {Sobczyk}},\
  }\href {\doibase 10.1103/PhysRevC.86.015505} {\bibfield  {journal} {\bibinfo
  {journal} {Phys. Rev. C}\ }\textbf {\bibinfo {volume} {86}},\ \bibinfo
  {pages} {015505} (\bibinfo {year} {2012})}\BibitemShut {NoStop}%
\bibitem [{\citenamefont {Buss}\ \emph {et~al.}(2012)\citenamefont {Buss},
  \citenamefont {Gaitanos}, \citenamefont {Gallmeister}, \citenamefont {van
  Hees}, \citenamefont {Kaskulov}, \citenamefont {Lalakulich}, \citenamefont
  {Larionov}, \citenamefont {Leitner}, \citenamefont {Weil},\ and\
  \citenamefont {Mosel}}]{GiBUU}%
  \BibitemOpen
  \bibfield  {author} {\bibinfo {author} {\bibfnamefont {O.}~\bibnamefont
  {Buss}}, \bibinfo {author} {\bibfnamefont {T.}~\bibnamefont {Gaitanos}},
  \bibinfo {author} {\bibfnamefont {K.}~\bibnamefont {Gallmeister}}, \bibinfo
  {author} {\bibfnamefont {H.}~\bibnamefont {van Hees}}, \bibinfo {author}
  {\bibfnamefont {M.}~\bibnamefont {Kaskulov}}, \bibinfo {author}
  {\bibfnamefont {O.}~\bibnamefont {Lalakulich}}, \bibinfo {author}
  {\bibfnamefont {A.}~\bibnamefont {Larionov}}, \bibinfo {author}
  {\bibfnamefont {T.}~\bibnamefont {Leitner}}, \bibinfo {author} {\bibfnamefont
  {J.}~\bibnamefont {Weil}}, \ and\ \bibinfo {author} {\bibfnamefont
  {U.}~\bibnamefont {Mosel}},\ }\href {\doibase
  https://doi.org/10.1016/j.physrep.2011.12.001} {\bibfield  {journal}
  {\bibinfo  {journal} {Physics Reports}\ }\textbf {\bibinfo {volume} {512}},\
  \bibinfo {pages} {1 } (\bibinfo {year} {2012})},\ \bibinfo {note}
  {transport-theoretical Description of Nuclear Reactions}\BibitemShut
  {NoStop}%
\bibitem [{\citenamefont {Oset}\ and\ \citenamefont {Salcedo}(1987)}]{OSMM}%
  \BibitemOpen
  \bibfield  {author} {\bibinfo {author} {\bibfnamefont {E.}~\bibnamefont
  {Oset}}\ and\ \bibinfo {author} {\bibfnamefont {L.~L.}\ \bibnamefont
  {Salcedo}},\ }\href {\doibase 10.1016/0375-9474(87)90185-0} {\bibfield
  {journal} {\bibinfo  {journal} {Nucl. Phys.}\ }\textbf {\bibinfo {volume}
  {A468}},\ \bibinfo {pages} {631} (\bibinfo {year} {1987})}\BibitemShut
  {NoStop}%
\bibitem [{\citenamefont {Abe}\ \emph {et~al.}(2015)\citenamefont {Abe},
  \citenamefont {Adam}, \citenamefont {Aihara}, \citenamefont {Akiri},
  \citenamefont {Andreopoulos}, \citenamefont {Aoki}, \citenamefont {Ariga},
  \citenamefont {Assylbekov}, \citenamefont {Autiero} \emph
  {et~al.}}]{T2K:flux2016}%
  \BibitemOpen
  \bibfield  {author} {\bibinfo {author} {\bibfnamefont {K.}~\bibnamefont
  {Abe}}, \bibinfo {author} {\bibfnamefont {J.}~\bibnamefont {Adam}}, \bibinfo
  {author} {\bibfnamefont {H.}~\bibnamefont {Aihara}}, \bibinfo {author}
  {\bibfnamefont {T.}~\bibnamefont {Akiri}}, \bibinfo {author} {\bibfnamefont
  {C.}~\bibnamefont {Andreopoulos}}, \bibinfo {author} {\bibfnamefont
  {S.}~\bibnamefont {Aoki}}, \bibinfo {author} {\bibfnamefont {A.}~\bibnamefont
  {Ariga}}, \bibinfo {author} {\bibfnamefont {S.}~\bibnamefont {Assylbekov}},
  \bibinfo {author} {\bibfnamefont {D.}~\bibnamefont {Autiero}},  \emph
  {et~al.},\ }\href {\doibase 10.1103/PhysRevD.91.072010} {\bibfield  {journal}
  {\bibinfo  {journal} {Phys. Rev. D}\ }\textbf {\bibinfo {volume} {91}},\
  \bibinfo {pages} {072010} (\bibinfo {year} {2015})}\BibitemShut {NoStop}%
\bibitem [{\citenamefont {Aguilar-Arevalo}\ \emph {et~al.}(2009)\citenamefont
  {Aguilar-Arevalo}, \citenamefont {Anderson}, \citenamefont {Bazarko},
  \citenamefont {Brice}, \citenamefont {Brown}, \citenamefont {Bugel} \emph
  {et~al.}}]{MBflux:2009}%
  \BibitemOpen
  \bibfield  {author} {\bibinfo {author} {\bibfnamefont {A.~A.}\ \bibnamefont
  {Aguilar-Arevalo}}, \bibinfo {author} {\bibfnamefont {C.~E.}\ \bibnamefont
  {Anderson}}, \bibinfo {author} {\bibfnamefont {A.~O.}\ \bibnamefont
  {Bazarko}}, \bibinfo {author} {\bibfnamefont {S.~J.}\ \bibnamefont {Brice}},
  \bibinfo {author} {\bibfnamefont {B.~C.}\ \bibnamefont {Brown}}, \bibinfo
  {author} {\bibfnamefont {L.}~\bibnamefont {Bugel}},  \emph {et~al.} (\bibinfo
  {collaboration} {MiniBooNE Collaboration}),\ }\href {\doibase
  10.1103/PhysRevD.79.072002} {\bibfield  {journal} {\bibinfo  {journal} {Phys.
  Rev. D}\ }\textbf {\bibinfo {volume} {79}},\ \bibinfo {pages} {072002}
  (\bibinfo {year} {2009})}\BibitemShut {NoStop}%
\bibitem [{\citenamefont {Aliaga}\ \emph {et~al.}(2016)\citenamefont {Aliaga},
  \citenamefont {Kordosky}, \citenamefont {Golan}, \citenamefont {Altinok},
  \citenamefont {Bellantoni}, \citenamefont {Bercellie},\ and\ \citenamefont
  {Betancourt}}]{NUMIflux}%
  \BibitemOpen
  \bibfield  {author} {\bibinfo {author} {\bibfnamefont {L.}~\bibnamefont
  {Aliaga}}, \bibinfo {author} {\bibfnamefont {M.}~\bibnamefont {Kordosky}},
  \bibinfo {author} {\bibfnamefont {T.}~\bibnamefont {Golan}}, \bibinfo
  {author} {\bibfnamefont {O.}~\bibnamefont {Altinok}}, \bibinfo {author}
  {\bibfnamefont {L.}~\bibnamefont {Bellantoni}}, \bibinfo {author}
  {\bibfnamefont {A.}~\bibnamefont {Bercellie}}, \ and\ \bibinfo {author}
  {\bibfnamefont {M.}~\bibnamefont {Betancourt}} (\bibinfo {collaboration}
  {MINER\ensuremath{\nu}A Collaboration}),\ }\href {\doibase
  10.1103/PhysRevD.94.092005} {\bibfield  {journal} {\bibinfo  {journal} {Phys.
  Rev. D}\ }\textbf {\bibinfo {volume} {94}},\ \bibinfo {pages} {092005}
  (\bibinfo {year} {2016})}\BibitemShut {NoStop}%
\bibitem [{NuW()}]{NuWroSITE}%
  \BibitemOpen
  \href@noop {} {\enquote {\bibinfo {title} {Nuwro user guide},}\ }\bibinfo
  {howpublished} {\url{https://nuwro.github.io/user-guide/}}\BibitemShut
  {NoStop}%
\bibitem [{\citenamefont {Adler}(1975)}]{Adler}%
  \BibitemOpen
  \bibfield  {author} {\bibinfo {author} {\bibfnamefont {S.~L.}\ \bibnamefont
  {Adler}},\ }\href {\doibase 10.1103/PhysRevD.12.2644} {\bibfield  {journal}
  {\bibinfo  {journal} {Phys. Rev. D}\ }\textbf {\bibinfo {volume} {12}},\
  \bibinfo {pages} {2644} (\bibinfo {year} {1975})}\BibitemShut {NoStop}%
\bibitem [{\citenamefont {Graczyk}\ \emph {et~al.}(2009)\citenamefont
  {Graczyk}, \citenamefont {Kie\l{}czewska}, \citenamefont {Przew\l{}ocki},\
  and\ \citenamefont {Sobczyk}}]{NuWroFF}%
  \BibitemOpen
  \bibfield  {author} {\bibinfo {author} {\bibfnamefont {K.~M.}\ \bibnamefont
  {Graczyk}}, \bibinfo {author} {\bibfnamefont {D.}~\bibnamefont
  {Kie\l{}czewska}}, \bibinfo {author} {\bibfnamefont {P.}~\bibnamefont
  {Przew\l{}ocki}}, \ and\ \bibinfo {author} {\bibfnamefont {J.~T.}\
  \bibnamefont {Sobczyk}},\ }\href {\doibase 10.1103/PhysRevD.80.093001}
  {\bibfield  {journal} {\bibinfo  {journal} {Phys. Rev. D}\ }\textbf {\bibinfo
  {volume} {80}},\ \bibinfo {pages} {093001} (\bibinfo {year}
  {2009})}\BibitemShut {NoStop}%
\bibitem [{\citenamefont {Nowak}\ and\ \citenamefont
  {Sobczyk}(2006)}]{NuWroDIS}%
  \BibitemOpen
  \bibfield  {author} {\bibinfo {author} {\bibfnamefont {J.~A.}\ \bibnamefont
  {Nowak}}\ and\ \bibinfo {author} {\bibfnamefont {J.~T.}\ \bibnamefont
  {Sobczyk}},\ }\href@noop {} {\bibfield  {journal} {\bibinfo  {journal} {Acta
  Phys. Polon.}\ }\textbf {\bibinfo {volume} {B37}},\ \bibinfo {pages} {2371}
  (\bibinfo {year} {2006})},\ \Eprint {http://arxiv.org/abs/hep-ph/0608108}
  {arXiv:hep-ph/0608108 [hep-ph]} \BibitemShut {NoStop}%
\bibitem [{\citenamefont {Bodek}\ and\ \citenamefont {Yang}(2002)}]{BODEKYANG}%
  \BibitemOpen
  \bibfield  {author} {\bibinfo {author} {\bibfnamefont {A.}~\bibnamefont
  {Bodek}}\ and\ \bibinfo {author} {\bibfnamefont {U.}~\bibnamefont {Yang}},\
  }\href {\doibase https://doi.org/10.1016/S0920-5632(02)01755-3} {\bibfield
  {journal} {\bibinfo  {journal} {Nuclear Physics B - Proceedings Supplements}\
  }\textbf {\bibinfo {volume} {112}},\ \bibinfo {pages} {70 } (\bibinfo {year}
  {2002})}\BibitemShut {NoStop}%
\bibitem [{\citenamefont {Sjostrand}\ \emph {et~al.}(2001)\citenamefont
  {Sjostrand}, \citenamefont {Lonnblad},\ and\ \citenamefont
  {Mrenna}}]{PYTHIA6}%
  \BibitemOpen
  \bibfield  {author} {\bibinfo {author} {\bibfnamefont {T.}~\bibnamefont
  {Sjostrand}}, \bibinfo {author} {\bibfnamefont {L.}~\bibnamefont {Lonnblad}},
  \ and\ \bibinfo {author} {\bibfnamefont {S.}~\bibnamefont {Mrenna}},\
  }\href@noop {} {\  (\bibinfo {year} {2001})},\ \Eprint
  {http://arxiv.org/abs/hep-ph/0108264} {arXiv:hep-ph/0108264 [hep-ph]}
  \BibitemShut {NoStop}%
\bibitem [{\citenamefont {Sobczyk}\ \emph {et~al.}(2005)\citenamefont
  {Sobczyk}, \citenamefont {Nowak},\ and\ \citenamefont {Graczyk}}]{WRONG}%
  \BibitemOpen
  \bibfield  {author} {\bibinfo {author} {\bibfnamefont {J.~T.}\ \bibnamefont
  {Sobczyk}}, \bibinfo {author} {\bibfnamefont {J.~A.}\ \bibnamefont {Nowak}},
  \ and\ \bibinfo {author} {\bibfnamefont {K.~M.}\ \bibnamefont {Graczyk}},\
  }\href {\doibase https://doi.org/10.1016/j.nuclphysbps.2004.11.218}
  {\bibfield  {journal} {\bibinfo  {journal} {Nuclear Physics B - Proceedings
  Supplements}\ }\textbf {\bibinfo {volume} {139}},\ \bibinfo {pages} {266 }
  (\bibinfo {year} {2005})},\ \bibinfo {note} {proceedings of the Third
  International Workshop on Neutrino-Nucleus Interactions in the Few-GeV
  Region}\BibitemShut {NoStop}%
\bibitem [{\citenamefont {Wilkinson}\ \emph {et~al.}(2014)\citenamefont
  {Wilkinson}, \citenamefont {Rodrigues}, \citenamefont {Cartwright},
  \citenamefont {Thompson},\ and\ \citenamefont {McFarland}}]{ANLBNLWilkinson}%
  \BibitemOpen
  \bibfield  {author} {\bibinfo {author} {\bibfnamefont {C.}~\bibnamefont
  {Wilkinson}}, \bibinfo {author} {\bibfnamefont {P.}~\bibnamefont
  {Rodrigues}}, \bibinfo {author} {\bibfnamefont {S.}~\bibnamefont
  {Cartwright}}, \bibinfo {author} {\bibfnamefont {L.}~\bibnamefont
  {Thompson}}, \ and\ \bibinfo {author} {\bibfnamefont {K.}~\bibnamefont
  {McFarland}},\ }\href {\doibase 10.1103/PhysRevD.90.112017} {\bibfield
  {journal} {\bibinfo  {journal} {Phys. Rev. D}\ }\textbf {\bibinfo {volume}
  {90}},\ \bibinfo {pages} {112017} (\bibinfo {year} {2014})}\BibitemShut
  {NoStop}%
\bibitem [{\citenamefont {Mosel}\ and\ \citenamefont
  {Gallmeister}(2017)}]{Mosel:T2KH2O}%
  \BibitemOpen
  \bibfield  {author} {\bibinfo {author} {\bibfnamefont {U.}~\bibnamefont
  {Mosel}}\ and\ \bibinfo {author} {\bibfnamefont {K.}~\bibnamefont
  {Gallmeister}},\ }\href {\doibase 10.1103/PhysRevC.96.015503} {\bibfield
  {journal} {\bibinfo  {journal} {Phys. Rev. C}\ }\textbf {\bibinfo {volume}
  {96}},\ \bibinfo {pages} {015503} (\bibinfo {year} {2017})}\BibitemShut
  {NoStop}%
\bibitem [{\citenamefont {Leitner}\ \emph {et~al.}(2009)\citenamefont
  {Leitner}, \citenamefont {Buss}, \citenamefont {Alvarez-Ruso},\ and\
  \citenamefont {Mosel}}]{Leitner:Resonances}%
  \BibitemOpen
  \bibfield  {author} {\bibinfo {author} {\bibfnamefont {T.}~\bibnamefont
  {Leitner}}, \bibinfo {author} {\bibfnamefont {O.}~\bibnamefont {Buss}},
  \bibinfo {author} {\bibfnamefont {L.}~\bibnamefont {Alvarez-Ruso}}, \ and\
  \bibinfo {author} {\bibfnamefont {U.}~\bibnamefont {Mosel}},\ }\href
  {\doibase 10.1103/PhysRevC.79.034601} {\bibfield  {journal} {\bibinfo
  {journal} {Phys. Rev. C}\ }\textbf {\bibinfo {volume} {79}},\ \bibinfo
  {pages} {034601} (\bibinfo {year} {2009})}\BibitemShut {NoStop}%
\end{thebibliography}%
